\documentclass[showpacs,amsmath,amssymb,twocolumn,superscriptaddress,notitlepage,preprintnumbers,pra,nofootinbib]{revtex4-2}
\usepackage{amsmath,amssymb,amsthm,mathrsfs,amsfonts,dsfont}
\usepackage{epsfig}
\usepackage{braket}
\usepackage{bm}
\usepackage{enumerate}
\usepackage{qcircuit}
\usepackage[svgnames]{xcolor}
\usepackage{graphicx}
\usepackage{comment}
\usepackage{algorithm}
\usepackage{algpseudocode}
\usepackage{balance}
\usepackage{tikz}
\usetikzlibrary{shapes,arrows}
\usepackage{verbatim}
\usepackage[normalem]{ulem}
\usepackage{varioref}

\usepackage{hyperref}
\hypersetup{
    colorlinks=true,
    linkcolor=Maroon,
    anchorcolor=Black,
    citecolor=DarkBlue,
    filecolor=blue,
    urlcolor=DarkBlue
}

\newcommand{\tr}{\operatorname{Tr}}
\newcommand{\RE}{\operatorname{Re}}
\newcommand{\IM}{\operatorname{Im}}

\labelformat{equation}{(#1)}

\labelformat{section}{#1}

\labelformat{figure}{#1}

\labelformat{proposition}{#1}

\labelformat{lemma}{#1}

\labelformat{theorem}{#1}

\labelformat{observation}{#1}

\labelformat{definition}{#1}

\labelformat{corollary}{#1}

\labelformat{problem}{#1}

\newcommand{\sun}[1]{\textcolor{black}{#1}}

\begin{document}

\title{Direct probing of the simulation complexity of open quantum many-body dynamics}

\date{\today}

\author{Lucia Vilchez-Estevez}
\email{lucia.vilchezestevez@physics.ox.ac.uk}
\affiliation{Clarendon Laboratory, University of Oxford, Parks Road, Oxford OX1 3PU, United Kingdom}

\author{Alexander Yosifov}
\email{alexanderyyosifov@gmail.com}
\affiliation{School of Physical and Chemical Sciences, Queen Mary University of London, London, E1 4NS, United Kingdom}

\author{Jinzhao Sun}
\email{Corresponding author: jinzhao.sun.phys@gmail.com}

\affiliation{Clarendon Laboratory, University of Oxford, Parks Road, Oxford OX1 3PU, United Kingdom}
\affiliation{School of Physical and Chemical Sciences, Queen Mary University of London, London, E1 4NS, United Kingdom}
 

\begin{abstract}

Simulating open quantum systems is key to understanding non-equilibrium processes, as persistent influence from the environment induces dissipation and can give rise to steady-state phase transitions. A common strategy is to embed the system-environment into a larger unitary framework, but this obscures the intrinsic complexity of the reduced system dynamics. Here, we investigate the computational complexity of simulating open quantum systems, focusing on two physically relevant parameters---correlation length and mixing time---and explore whether it can be comparable (or even lower) to that of simulating their closed counterparts. In particular, we study the role of dissipation in simulating open-system dynamics using both quantum and classical methods, where the classical complexity is characterised by the bond dimension and operator entanglement entropy. Our results show that dissipation affects correlation length and mixing time in distinct ways at intermediate and long timescales. Moreover, we observe numerically that in classical tensor network simulations, classical complexity does not decrease with stronger dissipation, revealing a separation between quantum and classical resource scaling.


\end{abstract}

\maketitle

\let\oldaddcontentsline\addcontentsline
\renewcommand{\addcontentsline}[3]{}


\section{Introduction}

In a way, quantum systems are in constant contact with a noisy environment \cite{breuer2002theory, rivas2012open}. Such openness gives rise to processes absent in closed systems, like decoherence and dissipation, which fundamentally shape the system dynamics and often drive irreversibility. Open quantum dynamics has therefore gained considerable attention due to its relevance in understanding non-equilibrium processes that are pervasive in real-world scenarios across biology \cite{mohseni2014quantum, mohseni2013geometrical}, materials science \cite{bauer2020quantum}, thermodynamics \cite{yunger2016microcanonical, rivas2020strong}, and control \cite{wiseman2009quantum, kallush2022controlling}. Simulating such open quantum systems is generally hard~\cite{weimer2021simulation}, as the system-environment interactions lead to nonunitary dynamics, driven by dissipative effects like relaxation and thermalisation \cite{cattaneo2023quantum, akemann2019universal, reichental2018thermalization, sa2021integrable}. A common way to deal with nonunitaries is by embedding into a unitary evolution (on an extended space), such as in Stinespring dilation \cite{hu2020quantum, ticozzi2017quantum,hu2022general}, purification \cite{schlimgen2022quantum}, unitary decomposition~\cite{schlimgen2022quantum_PRR,schlimgen2021quantum}, or in a more general context, by adding ancilla qubits~\cite{wang2011quantum,ding2024simulating,cleve2016efficient,li2023succinct,childs2016efficient,liu2021efficient,schlimgen2022quantum2}. 
In this way, the nonunitary dynamics of the reduced system is recovered by tracing out the environmental degrees of freedom. This idea has been instrumental in the algorithmic design of open quantum systems, enabling, for instance, the modelling of noisy quantum channels \cite{kretschmann2008information} and realisations based on a linear combination of unitaries or channels~\cite{cleve2016efficient,li2023succinct,childs2016efficient,liu2021efficient,an2023linear,an2023quantum,yu2024exponentially,kato2024exponentially}. 

Nevertheless, the embedding inherent to dilation often obscures the intrinsic complexity of the reduced system evolution~\cite{aloisio2023sampling}, making it hard to attribute observed complexity to the structure of the open quantum system itself, rather than to artifacts of the dilation procedure.
Meanwhile, the role of dissipation in the simulation complexity of open quantum dynamics has attracted interest on its own due to its nuanced nature. Dissipation can in fact lead to an increase in complexity in various scenarios, which is a bit counter-intuitive, as it is typically linked to the suppression of coherence and entanglement \cite{wang2024absence, alba2021spreading, zhao2025universal, barontini2013controlling}. As a concrete example, Prosen and Pizorn \cite{prosen2008quantum} showed that for an open Heisenberg XY spin chain, quantum phase transitions occur with small magnetic field, which exhibit long-range correlations in the non-equilibrium steady state and linear growth of the operator space entanglement entropy. Similarly, dephasing may lead to surprising behaviour at different time scales; specifically, the operator entanglement, while expected to be suppressed by dephasing, may rise, fall and rise again under time evolution~\cite{wellnitz2022rise}.
From a complexity point of view, Aloisio et al.~\cite{aloisio2023sampling} studied the complexity of sampling an open quantum system and demonstrated the existence of a stochastic quantum process that makes the multitime sampling as complex as sampling from a many-body state, which is classically hard. Therefore, a question arising from this is: Can we directly probe the complexity of an open quantum system and the contribution due to dissipation, via physically relevant parameters, without resorting to dilation. 

In this work, we study this question by exploring the complexity of simulating open quantum systems, focusing on physically meaningful parameters---\textit{correlation length} and \textit{mixing time}---in both short- and long-time regimes \cite{fang2025mixing, alipour2020correlation, babu2024unfolding, gribben2022using,kastoryano2013rapid}, as suggested in the context of quantum imaginary-time evolution (QITE) algorithms \cite{motta2020determining}. Specifically, we employ two complementary approaches: (i) classical tensor network (TN) simulations, and (ii) quantum simulation algorithms based on unitary approximation of nonunitary operators~\cite{motta2020determining} (see subsequent development in \cite{chan2023simulating,kamakari2022digital,sun2021quantum,jouzdani2022alternative,mao2023measurement,huang2023efficient}). We are particularly interested in how these spatial and temporal resources scale with respect to dissipation strength, time, and initial conditions of the system. Here, we consider a general open quantum system setup. We start with Markovian dynamics, including both time-dependent \cite{breuer2009stochastic} and nonlinear Lindbladian master equations \cite{liu2025digital} which can interpolate to non-Hermitian dynamics~\cite{liu2025lindbladian}. Then, we discuss the simulation of non-Markovian dynamics~\cite{tamascelli2018nonperturbative,menczel2024non,tanimura2020numerically,lambert2019modelling,park2024quasi,huang2025coupled,gao2022non,dan2025simulating,li2024toward,li2023dissipatons,yan2014theory,xu2022taming,xu2023universal}, in which we are focused on unravelling the influence functional of the environment with complex-valued paths proposed in \cite{xu2023universal} and simulation in an extended bosonic Hilbert space. To simulate the (non-)Markovian dynamics, we develop a stochastic simulation method based on quantum jump formulation~\cite{endo2020variational}. 
We analyse how dissipation affects the simulation complexity, where our technical contribution is linking it to the average correlation length, which is region- and time-dependent, and strictly smaller than the correlation length that is commonly used in literature (e.g., in \cite{motta2020determining}).

We show both analytically and numerically that increasing dissipation reduces the correlation length over time for quantum spin systems. This indicates that the dynamics of open quantum spins may be easier to simulate at long times than their closed counterpart. In the quantum case, we simulate the evolution of local spin observables in a dissipative spin chain with transversal and longitudinal fields up to 25 spins. The classical simulation complexity is characterised by the bond dimension and operator entanglement entropy, which quantify the growth of correlations and the cost of representing the system state efficiently. In contrast to the quantum case, we observe numerically that the complexity of classical TN simulations remains non-decreased with increasing dissipation strength, thereby showing a separation between quantum and classical scaling.
This suggests the existence of regions, where quantum methods may be easier to find advantages over classical techniques in dissipative chaotic spin systems. 
Finally, we consider an extension to ground-state preparation under time-dependent Lindbladians, where the jump operators are applied randomly at different sites over time in contrast to the dissipative methods in~\cite{lin2025dissipative,ding2024single}. 





\section{Results}
\label{sec:methods}

\subsection{Formulation of open quantum dynamics}
Many quantum processes can be captured by quantum Markov dynamics, where the evolution is recast into a quantum dynamical semigroup that generalises the Schr{\"o}dinger equation.  
We consider a quantum system $S$, described by Hamiltonian $H$ and density operator $\rho$, which are Hermitian in the operator space $\mathcal{A}(\mathcal{H})$ with Hilbert space $\mathcal{H}_{S}$; and suppose the evolution of $S$ is monitored by a detector with efficiency $\eta_{k} \in [0,1]$, see \cite{liu2025digital}. 

In particular, a spin system coupled linearly to a bosonic environment. 
The total Hamiltonian is defined on the composite Hilbert space 
\(\mathcal{H}_{\mathrm{tot}} = \mathcal{H}_S \otimes \mathcal{H}_B\),
where \(\mathcal{H}_B\) is the Hilbert space of the bosonic environment (or called bath). We consider both Markovian and non-Markovian regimes, where formally (in the weak coupling limit) the dynamics of $S$ is governed by a Lindblad master equation $\partial_t \rho = \mathcal{L}[\rho]$, with Lindbladian given by
\begin{equation}
\mathcal{L}[\rho] = -i[H, \rho] + \gamma \mathcal{N}[\rho] ,
\label{eq:timeL_non_linear}
\end{equation}
which consists of both unitary and dissipative components, denoted by $-i[H, \rho]$ and $\gamma \mathcal{N}[\rho]$, respectively, and $\gamma \geq 0$ is the interaction strength which dictates the rate of dissipation. The detector efficiency implies a nonlinear term in the Lindbladian. Explicitly
\begin{equation}
\begin{aligned}
\label{eq:lind_explicit}
\mathcal{N}[\rho] =&  \sum_k \alpha_{k}   \Bigl(
- \frac{1}{2} \{ L_k^\dagger L_k, \rho \} \\
&+ (1 - \eta_k ) L_k \rho L_k^\dagger +  \eta_k \braket{ L_k L_k^\dagger}_{\rho  } \rho \Bigr) ,
\end{aligned}
\end{equation}
where $\| \mathcal{N}\| = 1$. When $\eta_k= 0$, it reduces to the standard Lindbladian master equation, while for $\eta_k = 1$, it describes evolution under an effective non-Hermitian Hamiltonian~\cite{liu2025lindbladian}. The set of jump operators $\left\{L_{k}\right\} \subset \mathcal{A}(\mathcal{H})$ describe the dissipation in $S$, i.e., the decay of the off-diagonal terms of the positive semidefinite matrix $\rho$, while $\left\{\alpha_{k}\right\} \in \mathbb{R}$ are the corresponding decay coefficients; in the Markovian regime $\alpha_{k} > 0$. Here, each term in Eq.~\ref{eq:timeL_non_linear} is time-dependent and the coefficients are omitted for notational simplicity.
Unlike the linear case, this Lindbladian can capture non-Hermitian dynamics. From a computational perspective, its formulation does not allow for straightforward vectorisation.

This nonlinear process can be described in several ways. One way is to formalise it within a \textit{quantum instrument framework} as $\rho(t + dt) = \mathcal{E}(\rho(t))$, where
$ \mathcal{E} = \sum_k \mathcal{E}_{k}$ denotes a family of dynamical maps \cite{liu2025digital}.
Here, we consider an extended Hilbert space $S \otimes R$, where $R$ stores the classical record of the detector with orthogonal labels $\{ \ket{0}, |k, \textrm{jp}\rangle, |k, \textrm{disc}\rangle  \}_k$, denoting the three possible outcomes at step $k$: no jump, a jump takes place but remains undetected, and a jump is detected but is then discarded, respectively. Each of the three outcomes is associated with a corresponding operator acting on $S \otimes R$. Specifically
\begin{equation}
\begin{aligned}
&K_0 = |0\rangle\langle 0|_R 
\otimes \left( \mathbb{I} - i H\,\delta t 
             - \frac{1}{2} \sum_k \gamma \alpha_k L_k^{\dagger} L_k\,\delta t \right), \\
&K_{k, \mathrm{jp}} = |k, \mathrm{jp}\rangle\langle 0|_R 
\otimes \sqrt{\gamma \alpha_k(1-\eta_k) \,\delta t}\; L_k, \\[0.5em]
&K_{k, \mathrm{disc}} = |k, \mathrm{disc}\rangle\langle 0|_R 
\otimes \sqrt{\gamma \alpha_k \eta_k \delta t}\; L_k .
\end{aligned}
\end{equation}
It is easy to check that the reduced map on the system space $\Lambda (\rho) = \tr_R (\sum_k K_k (\rho \otimes |0\rangle \langle0| ) K_k^{\dagger})$ is completely positive trace-preserving (CPTP) in the first order of $\delta t$.
Given $S$ is in state $\rho$ at time $t$, the unnormalised post-selected state (labelled by $r \in \{0\} \cup \{(k, \textrm{jp})\}$) at $t+\delta t$ is 
\begin{equation}
\begin{aligned}
\rho^{\mathrm{un}}(t+\delta t) &= \rho 
+ \delta t \Big( -i[H,\rho] \\
&\quad + \gamma \sum_k  (1-\eta_k) \mathcal{D}_k[\rho] \Big) 
+ \mathcal{O}(\delta t^2), \nonumber
\end{aligned}
\label{eq:unnorm}
\end{equation}
where $\mathcal{D}_k[\rho] =  \alpha_k (L_k \rho L_k^\dagger - \frac{1}{2} \{ L_k^\dagger L_k, \rho \})$ is the dissipator. The corresponding success probability is then $p_{\text {keep }}=1-\sum_k \gamma \alpha_k \eta_k \delta t\langle L_k^{\dagger} L_k\rangle+\mathcal{O}(\delta t^2)$. Normalising $\rho^{\text{un}}$ yields the evolution given by Eq.~\ref{eq:lind_explicit}, where the last term is exactly the nonlinear drift that appears solely due to the renormalisation after discarding observed-click runs.

Another way to describe Eq.~\ref{eq:lind_explicit} makes use of the fact that it admits a natural interpretation in terms of quantum jumps \cite{gneiting2021jump, niu2023effect, perfetto2022thermodynamics}. We again use a detector with efficiency $\eta_k$ to monitor the occurrence of quantum jumps, where we suppose that a fraction of the data (given by $\eta_k$) is discarded as post-processing. The role of $\eta_k$ (and the post-selection) here is to interpolate the evolution between the normal dissipative and non-Hermitian dynamics. Considering the normalisation, a jump happens with probability $\frac{(1-\eta_k) \delta p_k}{ 1 - \eta_k \delta p_k} \approx (1- \eta_k) \delta p_k$, where $\delta p_k (t) := \delta t \alpha_k \tr( L_{k}^{\dagger} L_{k} \rho)$, whereas a ``no jump'' event occurs with probability $\frac{1- \delta p_k}{ 1 - \eta_k \delta p_k} \approx 1 - (1- \eta_k) \delta p_k$. From here, it is clear that this can be described by a quantum jump process. A system $S$ in pure state $\ket{\psi(t)}$ evolves for each small timestep $\delta t$ according to the stochastic dynamics: with probability $(1 - \eta_k) \delta p_k(t)$ the state jumps to $L_k\ket{\psi(t)} / {\|L_k \ket{\psi(t)} \|}$, while with probability $1 - (1 - \eta_k ) \delta p_k(t)$ it evolves to $e^{- \frac{1}{2} L_k^{\dagger} L_k \delta t } {\ket{\psi(t)}} / {\|e^{- \frac{1}{2} L_k^{\dagger} L_k \delta t } \ket{\psi(t)}\|}$, which leads to Eq.~\ref{eq:lind_explicit}  \cite{liu2025lindbladian}.

The above discussion uses a master equation to describe the dynamics of $S$. In general, the system-bath coupling induces finite environmental memory, and the dynamics cannot be captured by time-local Lindbladian operators alone. Let us consider the system-bath coupling $H_{\rm int} = {q}_S \otimes {q}_B$, with bath operator ${q}_B \in \mathcal{H}_{B}$ and a free Gaussian bath.  There are various approaches to unravel the influence of the bath, such as using pseudomodes~\cite{tamascelli2018nonperturbative,nusseler2022fingerprint,menczel2024non,zhou2024systematic,park2024quasi}, hierarchy of the equations of motion (HEOM)~\cite{tanimura2020numerically,xu2022taming,ke2022hierarchical}, dissipaton equation of motion~\cite{yan2014theory,yan2016dissipation}, dissipaton-embedded quantum master equation (DQME)~\cite{li2023dissipatons}, and  quantum dissipation with minimal state space (QD-MESS)~\cite{xu2023universal}. The simulation of non-Markovian dynamics on a quantum computer has also been discussed~\cite{li2024toward,guo2025variational,dan2025simulating}. A common approach is to characterise the environment by the bath correlation function $C^{\rm env}(t) = \braket{{q}_B(t) {q}_B (0)}$, and then decompose it into a finite number of complex exponential terms
\begin{equation}
C^{\rm env}(t)  = \sum_{k=1}^K d_k e^{-z_k t},    
\label{eq:c_env}
\end{equation}
where we choose to constrain the amplitude to be real $d_k \in \mathbb{R}^1$, while $z_k \in \mathbb{C}^1$ is a complex number. 
In this way, one can represent the dynamics of states in an extended space $\Gamma = \mathbb{L}_S \otimes \mathbb{F}_{2K}$, where $\mathbb{L}_S$ is the Liouville space of $S$, and $\mathbb{F}_{2K}$ is the Fock space of $2K$-mode harmonic system. 
In Supplementary Section~\ref{sec:appA}, we discuss the unravelling of the influence functional based on the extended state space introduced in \cite{xu2023universal} and how to map to a quasi-Lindblad equation. 

Formally, one can introduce a generator $\hat{\Gamma}_k = - z_k \hat{a}_k^{\dagger} \hat{a}_k - z_k^* \hat{b}_k^{\dagger} \hat{b}_k$ to describe the dynamics of the unperturbed auxiliary bosons in $\mathbb{F}_{2K}$, with $\hat{a}_k $ and $\hat{b}_k$ defined, respectively, as
\begin{equation}
\begin{aligned}
\hat{a}_k^{\dagger} \ket{\mathbf{m}} &= \sqrt{m_k + 1}\, \ket{\mathbf{n} + \mathbf{e}_k}, \\
\hat{b}_k^{\dagger} \ket{\mathbf{n}} &= \sqrt{n_k + 1}\, \ket{\mathbf{m} + \mathbf{e}_k},
\label{eq:ab}
\end{aligned}
\end{equation}
where $\mathbf{m} = \{m_1, ..., m_k,...,m_K\}$ and $\mathbf{n} = \{n_1, ..., n_k,...n_K\}$ label the HEOM $\rho_{\mathbf{m},\mathbf{n}}$. 
They are the quantum occupation numbers that are aligned with the multi-index in free-pole HEOM to label the auxiliary boson states, where $\ket{\mathbf{n}}$ denotes the pseudo-Fock states. In the main text,  the hat notation is used to denote the terms in the extended space. One may then apply a similarity transformation $\mathcal{S}$ with $\det(\mathcal{S}) = 1$ in Fock space that mixes $\hat{a}_k$ and $\hat{b}_k$ to map the time evolution of the density matrix to a Lindblad form when the amplitude is constrained to the real domain. 
Now to make it compatible for implementation on a quantum computer, one may further define the state by only using half of the auxiliary bosons that reside in a dual Fock space $ \{ \ket{\mathbf{m}}  \bra{\mathbf{n}} \}$. In doing so, it was shown in \cite{xu2023universal} that the dynamics is mapped to a Lindblad form
$\partial_t{\rho}_{\mathcal{S}} =  \hat{\mathcal{L}}_S [ \rho_\mathcal{S}]$, where
\begin{equation} 
\hat{\mathcal{L}}_S[\rho_{\mathcal{S}}]  =-i\left[\hat{H}_{\mathrm{pm}}, \rho_{\mathcal{S}}\right]+ \gamma \mathcal{D}_{\rm pm}[\rho_{\mathcal{S}}], 
\label{eq:lindd_pm}
\end{equation}
and the pseudomode Hamiltonian takes the form
\begin{equation}
\hat{H}_{\mathrm{pm}} =  {H} +\sum_k\left[\IM(z_k) \hat{a}_k^{\dagger} \hat{a}_k-\sqrt{\RE(d_k) } {q}_S \left(\hat{a}_k^{\dagger}+\hat{a}_k\right)\right],
\label{eq:H_eff_pm}
\end{equation}
with $\mathcal{D}_{\rm pm} = \sum_{k=1}^K \alpha_k \left[\hat{a}_k \rho_{\mathcal{S}} \hat{a}_k^{\dagger}-\frac{1}{2}\left\{\hat{a}_k^{\dagger} \hat{a}_k, \rho_{\mathcal{S}}\right\}\right]$, $\gamma = 2 \sum_{k=1}^K |\RE(z_k)|$, and $\alpha_k = 2 \RE(z_k) / \gamma$.
The dynamics of the state of $S$ is given by tracing out the degrees of freedom of the extended Hilbert space, $\rho(t) = \tr_{\backslash S} \rho_{\mathcal{S}}(t)$. This is similar to the pseudomode formulation of non-Markovian dynamics~\cite{lambert2019modelling,cirio2024modeling,luo2023quantum,nusseler2022fingerprint}, where the only difference compared to the Markovian case is that the dynamics of $S$ is defined on an extended space $\Gamma$ with auxiliary bosons characterised by $\hat{a}_k$. In
Supplementary Section~\ref{sec:appA}, we show how to get the Lindblad dynamics from the auxiliary paths defined on the product space. 
Compared to the formulation by DQME~\cite{li2024toward}, the dissipative operators and density operators are defined upon the Hilbert space $\mathcal{H}_S   \otimes \mathcal{H}_F   $. If the amplitude is not constrained to be real, then the reduced system dynamics is described by a quasi-Lindblad master equation defined on the extended space. In this case, the vectorised form of the generator $ \hat{\mathcal{L}}_S$ can still be written as 
$ 
{\mathcal{L}}_S = -i (H_1 - i H_2),
$
where $H_1$ and $H_2$ denote, respectively, the Hermitian and anti-Hermitian matrices, see Eq.~\ref{eq:NM_general_DM} in Appendix.

\subsection{Simulating the open-system dynamics}

From the above discussion, simulating open-system dynamics can be mapped onto the simulation of a Lindbladian (or quasi-Lindbladian) evolution.
Both Markovian and non-Markovian dynamics take a form similar to Eq. \ref{eq:lind_explicit} with the latter requiring an enlarged (dilated) space.
Such dynamics can be simulated either by directly evolving the dilated system or through stochastic approaches based on pure-state trajectories.
Here, we mainly focus on the latter. To that end, we consider the stochastic Schr\"odinger equation's approach to simulate the Lindblad equation. To do so, one may employ either a discretised or a continuous method. The circuit complexity in the discretised approach inevitably depends on the number of steps. To avoid this, we present a stochastic Monte Carlo sampling method. The algorithm is described below.

The jump time is computed classically based on the total jump probability (related to $\delta p_k$ and $\eta_k$) using the sampling method in \cite{endo2020variational}, where the key task is to compute the probability that there will be \textit{no jump} in a given time period.
The total jump probability in the short duration from $t$ to $t+\delta {t}$ is $P_{\textrm{jp}}(t) = {\gamma}  \sum_k (1 - \eta_k ) \delta p_k (t)$. Given the state undergoes a quantum jump at time $t$, the probability that there will be no jump until $t + \tau$ is $$Q(t + \tau) = \lim_{\delta {t} \rightarrow 0} \prod_{j}^{\tau/\delta {t}} (1 - P_{\textrm{jp}}(t +  j \delta {t} ) ) = e^{- \int_{t}^{ t + \tau} P_{\textrm{jp}}(\tau') d {\tau}'}.$$
From this, we can compute the sequence of jump times ${t_i}$, which in turn determines the ensemble of trajectories; specifically, one can generate a random number $q \in [0,1]$. Solve  $\tau = \mathrm{argmax }_{\tau} ( Q(t + \tau)  \leq q)$, which is equivalent to finding the largest $\tau$ such that $\int_{t}^{ t + \tau} P_{\textrm{jp}}(\tau') d {\tau}' \leq - \ln(q)$.

In case there is no jump in a given time period, the state continues to evolve under the effective operator
\begin{equation}
\label{eq:Heff}
H_{\text{eff}} =  H + \frac{{i \gamma}}{2}\sum_{k}\left(L_k^\dagger L_k -  \braket{L_k^\dagger L_k }\right),
\end{equation}
where the subtraction of $\braket{L_k^\dagger L_k }$ is due to the normalisation after applying $e^{- \frac{1}{2} L_k^{\dagger} L_k \delta t}$. 
In the case of Markovian dynamics, $H = H_S$ is the system Hamiltonian and $L_k$ is introduced in Eq. \ref{eq:lind_explicit}. For non-Markovian dynamics, $H = \hat{H}_{\mathrm{pm}}$ and $L_k = \alpha_k \hat{a}_k$ in Eq. \ref{eq:lindd_pm}. 
For each trajectory, we start from an initial state $\ket{\psi(0)}$ which we evolve under Eq.~\ref{eq:Heff} by using the QITE algorithm \cite{motta2020determining} until the jump time. 
This process is repeated until the final simulation time $T$ is reached, generating a single trajectory. Observable quantities are then estimated by averaging over many such trajectories.


The complexity of the simulation is mainly determined by the cost of simulating the non-Hermitian operator $H_{\text{eff}}$ and quantum jumps on a digital quantum computer.
Quantum jumps interrupt the dynamics only through interleaving with the jump operator $L_k$, which has a singular value decomposition $L_k = U_{k} D_k V_{k}$, with unitaries $U_k$ and $V_k$, and a diagonal operator $D_k = \sum_j a_{k, j} \ket{j}\bra{j}$. As shown in \cite{endo2020variational}, $D_k$ can be realised by a nonunitary exponentiated operator $D_k \approx   \exp( -H^{D_k} T^{D_k})$, with $-H^{D_k} T^{D_k} = \sum_{a_{k, j} \neq 0 } \log(a_{k, j} ) \ket{j} \bra{j} - b_{D_k}  \sum_{a_{k, j} = 0 } \ket{j} \bra{j} $. Here, $b_k$ is an appropriately chosen constant, determined by $\braket{D_k^2}$, that bounds the approximation error, see~Supplementary Section~\ref{app:totalerror} for details and examples. As $L_k$ usually acts locally, $H^{D_k}$ is also local, making it efficient to approximate $D_k$ by the QITE algorithm. 

We remark that the average number of jump events is  $\mathcal{O} (\gamma  T \sum_k  \alpha_k)$, which is linear in $T$.
This realisation of the jump integrates naturally with the QITE framework. Therefore, in what follows, we primarily consider the cost associated with simulating $H_{\text{eff}}$ and consider a continuous evolution up to time~$T$.

\emph{Sketch of the quantum simulation procedure}.
From the above discussion, it is easy to see that the Markovian \ref{eq:timeL_non_linear} and non-Markovian dynamics \ref{eq:lindd_pm} can be simulated using the stochastic method. The procedure is as follows. Evolve the state under $H_{\rm eff}$ until the next jump happens. The probability that the next jump happens at time $t + \tau$ is $Q(t + \tau)  $, based on which the jump time $\tau$ can be determined. Then, we update the quantum state by applying the quantum jump $L_k$ which is selected according to the weight $\alpha_k$.
 
For non-Markovian dynamics with pseudomode decomposition, given the bath correlation function, truncate pseudomodes to $K$ with which to construct the effective Hamiltonians \(\hat H_{\mathrm{eff}}(t) \in \mathcal{H}_S \otimes \mathcal{H}_F\). If we choose to constrain $d_k$ to be real, then we can use the stochastic method with pure states only to simulate the dynamics $\hat{L}_S$.  In practice, we truncate the energy level of the pseudo-Fock space to $N_B$ levels. Then, the total number of qubits is $N + K \log(N_B)$.
Encode the initial state on a quantum computer.
In the general case where $\IM(d_k) \neq 0$, the dynamics is given by Eq. \ref{eq:NM_general_DM}. In this situation, one can perform vectorisation to simulate the Hermitian and anti-Hermitian components in $\hat{L}_S$, described in Eq. \ref{eq:NM_vector}. The total number of qubits is $2(N + K \log(N_B))$. The rest of the procedure is thus similar to that for simulating the linear Lindblad master equation.

\emph{Remarks on the role of dissipation.}
QITE-based simulations of dissipation typically enlarge the effective support of the evolution operator, suggesting a higher complexity compared to purely unitary dynamics. Still, this increase is not universal. In the case of strong dissipation---where the dissipative term acts effectively like a projector---the complexity is determined mainly by the projector’s support and becomes independent of the total evolution time. Consequently, for a fixed $t$, increasing $\gamma$ can, in fact, reduce the required support, and thereby lower the QITE simulation complexity, Fig. \ref{fig:correlations_chaos}. Moreover, even for weaker dissipation, its effects accumulate over time and can eventually dominate the dynamics, again leading to reduced complexity. This supports our conjecture that there exists a timescale beyond which the complexity of simulating open quantum systems is below that of Hamiltonian systems. Yet, at very short times, the situation is more subtle: the use of a unitary approximation to represent the nonunitary evolution can temporarily increase the complexity.

To enable a fair comparison between open quantum system and unitary dynamics in later section, we introduce the operator $\tilde{H}_{\text{eff}} = H_{\text{eff}}/\| H_{\text{eff}}\|$, i.e., a normalised version of Eq.~\ref{eq:Heff}. Explicitly, we divide Eq.~\ref{eq:Heff} into the Hermitian and non-Hermitian part, $\tilde{H}_{\text{eff}} = \lambda_1 H_1 - i \lambda_2 H_2$, where $\lambda_1^2+\lambda_2^2 = 1$, and $\|H_1\| = \|H_2\| = 1$.
\sun{By normalising the effective operator, the final time can be regarded as stretched and we denote this rescaled final time by $T$.}
Overall, the procedure involves decomposing $H_{\text{eff}}$ into $H_1$ and $H_2$, allowing the representation of the time-evolution operator as a product of exponentials $\ket{\psi(t)} = \exp[-i(\lambda_1 H_1 t - i\lambda_2 H_2 t)] \ket{\psi(0)}$. Using Trotterisation, the nonunitary dynamics is approximated by an alternating sequence of unitary operations, generated by $H_1$, and QITE-based steps that incorporate the effects of $H_2$ as $\ket{\psi(t)} = [\exp(-i \lambda_1  H_1 \delta {t} ) \exp(- \lambda_2  H_2 \delta {t} )]^{T /  \delta {t}}\ket{\psi(0)} + \mathcal{O}(T \delta {t} )$. Here a first-order Trotter decomposition is applied as an example, while $K$th-order Trotter decomposition is used in deriving the analytical bound in Supplementary Section~\ref{app:totalerror}. The first part $H_{1}$ can be implemented on a quantum computer by applying standard simulation techniques for Trotterisation \cite{han2021experimental, pocrnic2023quantum}. In contrast, the nonunitary component $H_{2}$ cannot be directly realised on quantum hardware and is instead simulated using QITE proposed in \cite{motta2020determining}.

Before we continue, we ought to compare our approach with others in the literature. There are two key aspects that distinguish the above continuous stochastic method from that of the discretised approach \cite{liu2025digital}. First, within the QITE framework, the described quantum jump process enables both nonlinear and linear Lindbladians in Eq.~\ref{eq:Heff} to be simulated deterministically, in contrast to the probabilistic realisation of the nonlinear Lindbladian simulation with two ancillary qubits per timestep \cite{liu2025digital}.
Second, this approach does not explicitly involve Trotterisation but rather takes a stochastic, random-sampling approach that recovers the ideal trajectory on average. It is thereby compatible with digital-analogue types of quantum simulation, which may not necessarily incur the additional cost associated with Trotterisation. Moreover, the above method directly applies to time-dependent Lindbladians, as we show in Fig. \ref{fig:combined_lindblad} of the Supplementary Section \ref{sec:app_time_dep}. 

Finally, we comment on the possible strategies for simulating the Lindbladian master equation by density matrix vectorisation, where the density matrix is mapped to a pure state in a Liouville space~\cite{an2023quantum}. The dynamics is then simulated by applying the Lindbladian superoperator to this {vectorised} state \cite{gilchrist2009vectorization}. As with the generalised evolution operator, the Lindbladian can also be decomposed into normalised Hermitian and non-Hermitian parts. Here, one can find that the vectorised Lindbladian in \cite{kamakari2022digital} takes a similar form to Eq.~\ref{eq:Heff}, which can be divided into Hermitian and non-Hermitian parts, but acting on two copies of the state. Our stochastic approach is different from the methods for open quantum systems proposed in \cite{chan2023simulating} and \cite{kamakari2022digital}, where the vectorised form of the system's density matrix $\rho$ is encoded in a larger state vector. A key consequence is that they can only preserve $\tr(\rho^2)$, rather than $\tr(\rho)$. As a result, the observable expectation value is calculated by $\braket{O} / \tr(\rho)$, with $\tr(\rho)$ being measured at each timestep.
Note that the complexity analysis of the full evolution of the vectorised density matrix remains the same for the pure-state evolution, as it adopts a similar form with unitary and nonunitary components, see Supplementary Section~\ref{sec:app_vector_density}.

\sun{For non-Markovian dynamics, the formulation given by Eq.~\ref{eq:lindd_pm} is norm-conserving. In contrast,  in DQME-SQ the state dynamics are non-Hermitian and not trace-preserving. The norm of the state may be decreased with both time $t$ and the number of dissipatons in the extended space, due to the non-Hermitian term of $\hat a_k^{\dagger}$ alone (instead of $\hat a_k^{\dagger} + \hat a_k$).  Therefore, the resulting states carry a failure probability after post-selection, which will introduce additional sample complexity when implementing on a quantum computer.
}


\subsection{Physical measures of quantum algorithmic complexity}
\label{sec:complexity}

We now wish to analyse the quantum algorithmic complexity in terms of physically interpretable quantities.
It is well known that in the short-time regime, open quantum systems exhibit growing spatial correlations due to information propagation, which influences the computational resources required to simulate them \cite{poulin2010lieb, schuch2008entropy}.\footnote{In closed many‑body systems, this growth is constrained by the Lieb–Robinson bounds \cite{lieb1972finite}.} For a quantum system $S$ in state $\ket{\Psi_S}$, usually the following definition is adopted (see \cite{brandao2015exponential,brandao2019finite}):
we say that $\ket{\Psi_S}$ exhibits $\xi$-exponential decay of correlations if there exists a correlation length $\xi > 0$ such that for all pairs of disjoint regions 
$(A,B)$ with separation $d(A,B) \ge l_{0}$, where $l_{0} \in \mathbb{R}^{+}$ is a fixed 
distance threshold, and for all operators $\hat{A} \in \mathcal{A}(\mathcal{H}_A)$, 
$\hat{B} \in \mathcal{A}(\mathcal{H}_B)$, the bound
\begin{equation*}
\label{eq:cor}
    \mathrm{Cor}(\hat{A},\hat{B}) 
    \le \|\hat{A}\| \, \|\hat{B}\| \, e^{-\frac{d}{\xi}}
\end{equation*}
holds.
Here, $\xi$ corresponds to the slowest decaying correlation in $S$ across all region pairs and all times \cite{hastings2006spectral}, and the correlation function reads
\begin{equation}
    \mathrm{Cor}(\hat{A},\hat{B}) :=
    \langle \Psi_S | \hat{A} \otimes \hat{B} | \Psi_S \rangle 
    - \langle \Psi_S | \hat{A} \otimes \mathbb{I} | \Psi_S \rangle 
      \langle \Psi_S | \mathbb{I} \otimes \hat{B} | \Psi_S \rangle ,\nonumber
\end{equation}
where for a pair $(A,B)$ at distance $d$, one can always define a correlation length $C_{A,B} := -\frac{d(A,B)}{\log\left(\frac{|\mathrm{Cor}(A,B)|}{\| {A}\| \| {B}\|}\right)} \leq \xi$ that satisfies the definition of correlation length for a given choice of regions. This bound applies for a fixed state $\ket{\Psi_S}$.

In this work, we wish to establish a tighter bound that can track correlation decay during the time evolution of $S$. Particularly, we consider a pair-specific, distance- and time-dependent correlation length as a replacement of $\xi$. More formally, we take $C_{A,B}(t)$ to be the infimum over all such $ \mathrm{Cor}(\hat{A},\hat{B})$
\begin{equation}
\label{eq: time_dep_corr}
    C_{A,B}(t) := \inf \left\{C > 0:
    \left| \mathrm{Cor}(\hat{A},\hat{B}) \right| 
    \le \|\hat{A}\| \, \|\hat{B}\| \, e^{-\frac{d}{C}}\right\},
\end{equation}
where $C$ controls how fast the correlations between $A$ and $B$ decay with $d$. From all pair-specific correlation lengths, we can thereby recover $\xi$ as the worst-case across all pairs of regions $\xi(t) := \sup_{A,B} C_{A,B}(t)$.

To define the average of Eq.~\ref{eq: time_dep_corr}, let $\mathcal{P}_{d}$ denote the set of all pairs of regions $(A,B)$ at distance $d$, then
\begin{equation}
\label{eq: averaged}
    \overline{C(d,t)} := \frac{1}{|\mathcal{P}_d|} 
    \sum_{(A,B) \in \mathcal{P}_d} C_{A,B}(t).
\end{equation}
From the concavity of the function $f(x) = e^{-d/x}$ when $x \geq {1}/{2}$, and using Jensen’s inequality, we find that $$\overline{ \mathrm{Cor}(d,t)} \leq \overline{ e^{-d/C_{A_i, B_i}}} |_ {A_i - B_i = d} \leq e^{-d/ \bar{C}}$$ when $\min_i C_i \geq 1/2.$ The average correlation length is introduced to capture the general case when $C_{A,B}(t)$ varies across different pairs of regions. The correlation length $\xi(t)$ captures the worst-case behaviour, whereas the average behaviour is captured by $C_{A,B}(t) \leq \xi(t)$, which gives a tighter bound for that pair as it is defined by the smallest value of $C$ that satisfies the inequality \ref{eq: time_dep_corr}. 
These quantities will later be used to relate the error bounds and algorithmic complexity. As we shall see later, the algorithmic complexity is mainly determined by the average correlation length $\overline{C}$. In Supplementary Section \ref{app:totalerror}, we describe its scaling, including Trotter errors and the dependence of the total simulation error on $\overline{C}$. 



The above discussion focuses on methods that operate within a fixed Hilbert space. Before we move to the numerical simulation, we would like to comment on how the complexity scales when we allow for systematic expansion via additional ancilla qubits.
Due to locality and finite correlation length, each of the $N$ sites in $S$ is influenced primarily by bath modes within a spatial radius $\xi$. Therefore, for each site per each timestep, we only need to keep track of the part of the environment memory within distance $\xi$. Putting all together, the total complexity (i.e., ancilla $\times$ circuit depth) is $N m \xi N_{\text{steps}} = \mathcal{O} (N \xi (\lambda T)^{ 1+ o(1)}/\varepsilon^{o(1)})$, which is polynomial in correlation length and simulation time, where $N_{\text{steps}}$ is the number of Trotter steps. This may support the conjecture that under reasonable physical assumptions, the cost of simulating an open quantum system with a finite correlation length is efficiently bounded. We refer to Supplementary Section \ref{app:spacetime} for details. On the other hand, in the asymptotic long-time limit, $S$ approaches a unique steady state $\rho_\infty$. Where the rate of convergence is controlled by the spectral gap $\Delta$ of the Liouvillian $\mathcal{L}$, and the mixing time satisfies:
$t_{\mathrm{mixing}} \sim \Delta^{-1}.$ After $t_{\mathrm{mixing}}$, the quantum simulation no longer requires evolving time-dependent correlations. The required ancilla size saturates to $C_{\mathrm{mixing}}$, and the number of steps becomes independent of the initial state. Thus, the quantum complexity is saturated up to a certain timescale. 
\subsection{Investigation on quantum and classical runtime}

\noindent
\textbf{Quantum runtime analysis.}
In this section, we study the influence of dissipation on two physically relevant parameters in both the short- and long-time regimes, which will inform us about the quantum algorithmic complexity of simulating the system of interest $S$. In particular, we will focus on the \textit{mixing time}, which characterises how quickly $S$ reaches its steady state, and the \textit{correlation length}, which quantifies the spatial extent of quantum correlations within $S$. Both provide physical insight into the dissipative dynamics and serve as key indicators for assessing the simulability of $S$. We numerically analyse the dependence of simulation complexity on the dissipation strength $\gamma$, evolution time $t$, and system size $N$.

\begin{figure}[!t]
    \centering
    \includegraphics[width=\linewidth]{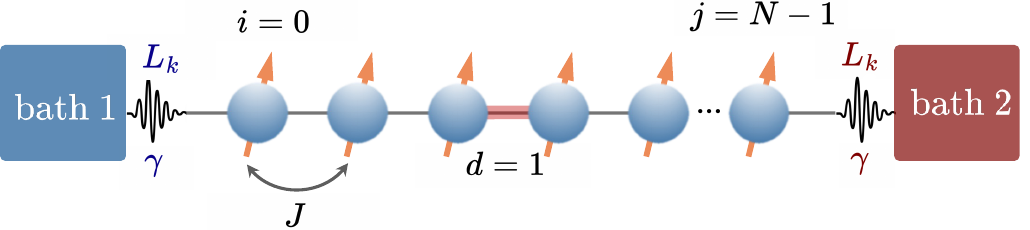}
    \caption{\textbf{Illustration of the dissipative spin system under study.} The boundary-driven quantum spin chain $S$ is coupled with strength $\gamma$ to two baths at its edges, $i=0$ and $j=N-1$, and the set of jump operators $\left\{L_{k}\right\}$ acts locally at the boundary sites. Spatial separation of $d=1$ means two sites are nearest neighbours with coupling $J$.}
    \label{fig:spin_chain}
\end{figure}

We begin by describing our setup. We consider a quantum spin chain system $S$ with $N$ sites, described by a one-dimensional Ising Hamiltonian with transverse and longitudinal fields
\begin{equation}
    H_{I} = J\sum_{i} X_i X_{i+1} + h_x\sum_iX_i + h_z\sum_i Z_i,
    \label{eq:ising}
\end{equation}
where $J$ is the nearest-neighbour coupling strength, and $h_x$ and $h_z$ denote the magnetic fields in the $X$ and $Z$ directions, respectively. We set $J = 1$ and choose $h_x = h_z = -2$, as in this regime the expectation values of the local operators typically exhibit oscillatory behaviour, making the model suitable for testing our conjecture about the Lindbladian evolution. 
Here, $S$ is in contact with an external environment, as described by the time-independent version of Eq.~\ref{eq:timeL_non_linear}, see Fig. \ref{fig:spin_chain}. We couple $S$ to two baths (modelled as local Markovian reservoirs) at its two edges, $i=0$ and $j=N-1$, and make a simple but common choice

\begin{equation}
\begin{aligned}
L_1 &= \sigma^+_{0}, &\quad \alpha_1 &= \gamma (1 + \mu), \\
L_2 &= L_1^\dagger = \sigma^-_{0}, &\quad \alpha_2 &= \gamma (1 - \mu), \\
L_3 &= \sigma^+_{N-1}, &\quad \alpha_3 &= \gamma (1 - \mu), \\
L_4 &= L_3^\dagger = \sigma^-_{N-1}, &\quad \alpha_4 &= \gamma (1 + \mu),
\end{aligned}
\nonumber
\end{equation}
where $\gamma$ is the system-bath coupling strength which determines the dissipation rate, $\mu$ is the driving parameter, and $\sigma^{\pm} = \sigma^x \pm i\sigma^y$. In Supplementary Section \ref{sec:app_tensor_network} we present additional results with $h_x=0.5$ and $h_z=-1.05$, corresponding to a chaotic regime, see Fig. \ref{fig:pauli_expectation_chaos}. The correlation length there exhibits the same trend, becoming increasingly damped with growing dissipation strength and total evolution time, but more noticeably than in the case studied in the main text.

We fix the driving part of the Lindbladian and are primarily interested in how $\gamma$ impacts the dynamics of $S$. The jump operators $\left\{L_{k}\right\}$ are assumed to act locally at the boundaries of the spin chain, leaving the coherent Hamiltonian-driven evolution in the bulk unaffected. Such boundary-driven setups are conceptually appealing as they are analogous to classical nonequilibrium lattice models. While a rigorous microscopic derivation of these local operators from weak system-bath coupling typically yields nonlocal terms, this approach remains physically insightful. It captures key features of non-equilibrium steady states, quantum transport, and dissipation-induced dynamics. This makes the framework well-motivated for exploring questions in quantum thermodynamics and boundary-driven open quantum systems~\cite{benatti2021exact, landi2022nonequilibrium}. In the simplest case of single-site boundary driving, one typically introduces two Lindblad operators that flip a spin up or down with different probabilities, thereby inducing a net magnetisation at the boundary site.

To simulate the open quantum system dynamics, the vectorised density matrix is encoded as a matrix product state (MPS) with physical dimension $4$ for each of the $N$ sites, \sun{corresponding to a state in the Liouville space of dimension $4^N$, see Supplementary Section \ref{sec:app_vector_density}}. The Liouvillian superoperator ${\mathcal{L}}$ is represented as a matrix product operator (MPO) with the same enlarged dimension. Time evolution is carried out using the time-evolving block decimation (TEBD) algorithm, where the exponential $e^{{\mathcal{L}}\delta t}$ is approximated via a second-order Trotter-Suzuki decomposition into sequentially applied local exponentials. The timestep $\delta t$ is dynamically adjusted to ensure consistent numerical accuracy throughout the simulation. The classical dependence is related to the bond dimension $\chi$. To ensure the simulation accuracy, we keep track of $\chi$ and check if the error converges with increasing $\chi$.

\begin{figure}[t]
    \centering
    \includegraphics[width=1.0\linewidth]{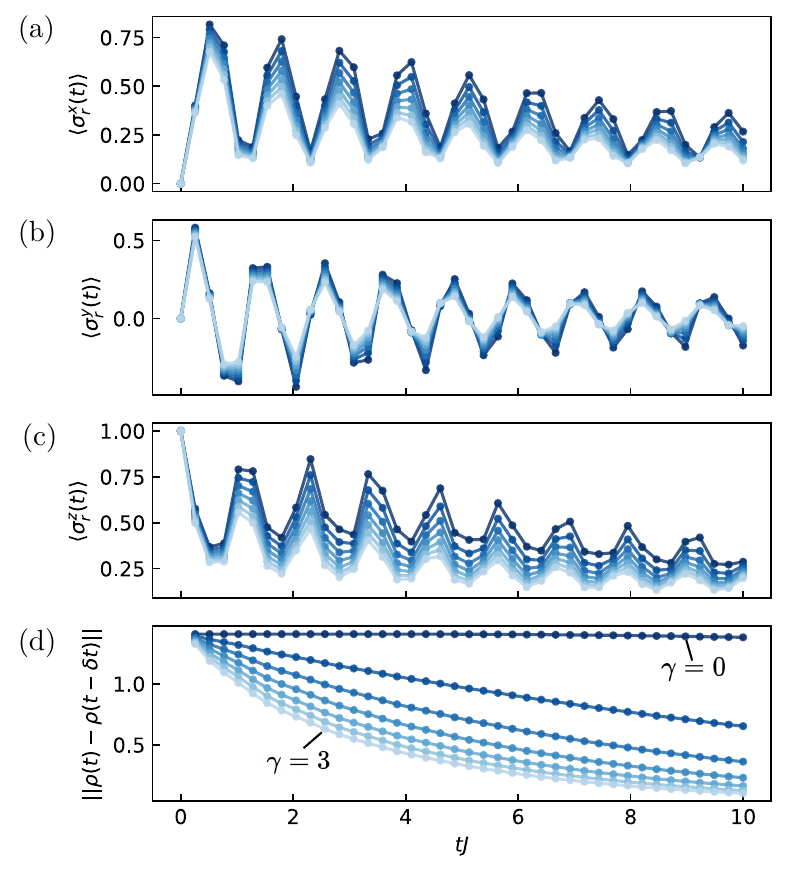}
    \caption{\textbf{Time evolution of spin observables and trace distance for a dissipative boundary-driven spin chain system $S$ of size $N = 25$ under increasing dissipation rates $\gamma$.} (a–c) Expectation values of the local spin operators $\langle \sigma^x_r(t) \rangle$, $\langle \sigma^y_r(t) \rangle$, and $\langle \sigma^z_r(t) \rangle$, respectively, under Lindbladian dynamics for different values of $\gamma$. (d) Trace distance between consecutive timesteps, $D = ||\rho(t) - \rho(t - \delta t)||$, indicating convergence toward a steady state. In all panels, $\gamma \in [0, 3]$, with step size of $0.05$, where increasing values are shown in progressively lighter shades of blue. Simulations were performed using MPO methods with $\chi = 250$.}
    \label{fig:pauli_expectation}
\end{figure}

Before analysing global quantities, it is useful to first examine how dissipation manifests locally in $S$. Local \textit{magnetisation} observables offer a direct window into the interplay between coherent dynamics and environmental noise: they reveal how quickly local excitations lose their quantum character and relax towards equilibrium. By studying their time evolution for different values of $\gamma$, we can build physical intuition for the mechanisms that ultimately determine the simulability of $S$.
Here, for each site $r$ and spin component $a \in \{X,Y,Z\}$, the expectation value is given by
$ 
    \langle \sigma^a_r (t)\rangle = \tr [\sigma^a_r \rho(t)],
$
where $\rho(t)$ is the density matrix evolved under Eq.~\ref{eq:timeL_non_linear} with Eq.~\ref{eq:ising} as the Hamiltonian. Fig. \ref{fig:pauli_expectation}(a)-(c) show the time evolution of the expectation values of these local Pauli operators for various $\gamma$. As expected, increasing $\gamma$ results in stronger damping of the oscillations, reflecting how stronger environmental coupling suppresses coherent dynamics and drives $S$ towards a steady state at long times.
In the closed case, oscillations persist over long times, reflecting coherent spin exchange in the Heisenberg model. As $\gamma$ increases, these oscillations decay more rapidly and their amplitude is strongly suppressed, signalling the progressive loss of phase coherence and the dominance of relaxation processes. Physically, this means that information about the initial state is erased faster, $S$ approaches a featureless steady state sooner, and the dynamical complexity is reduced. In the broader context of simulability, such damping hints at shorter correlation lengths and mixing times, which can make long-time dynamics more accessible to classical methods and reduce the potential advantage of quantum algorithms in this regime.

The damping of local magnetisation oscillations observed in Fig.~\ref{fig:pauli_expectation}(a)–(c) already suggests that dissipation accelerates the relaxation of $S$ towards equilibrium. To quantify this effect in a more global and model-independent way, we turn to the \textit{mixing time}---the timescale over which $S$ effectively reaches a steady state. This provides a direct indicator of how long coherent or correlated dynamics persist, and therefore how challenging the long-time regime may be to simulate.
We quantify this by computing the trace distance $D(t) = \|\rho(t) - \rho(t-\delta t)\|$ between density matrices at successive timesteps for various values of $\gamma$. Defining a small threshold $\varepsilon$, $S$ is considered to have effectively reached its steady state when $D < \varepsilon$. The results in Fig. \ref{fig:pauli_expectation}(d) show that stronger dissipation leads to faster convergence, i.e., shorter mixing time, consistent with the relaxation of local observables in Fig. \ref{fig:pauli_expectation}(a)-(c).
\begin{figure}[t]
    \centering
    \includegraphics[width=1.0\linewidth]{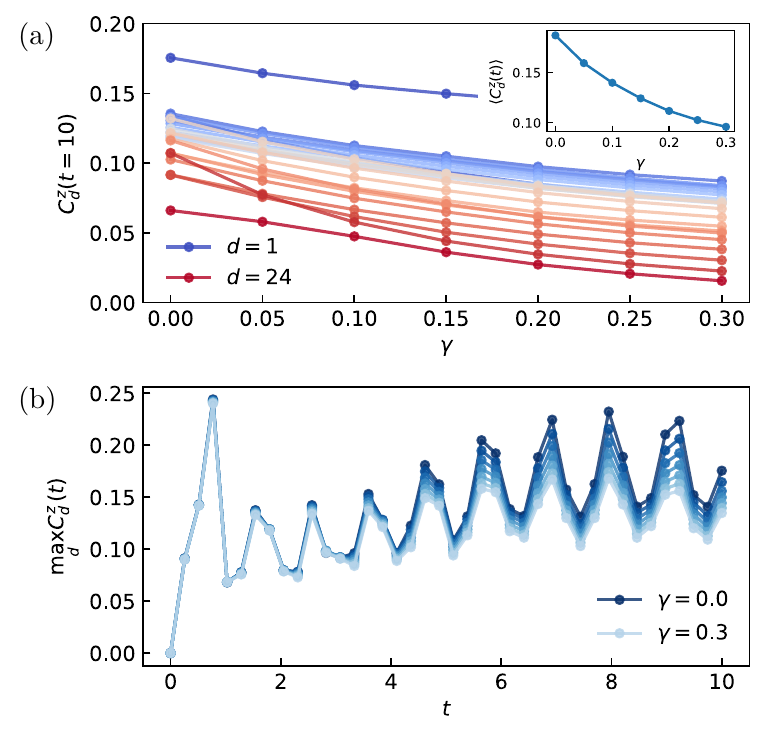}
    \caption{\textbf{Evolution of spin-$Z$ correlations in a dissipative boundary-driven spin chain $S$ with $N=25$ sites as a function of $\gamma$, $t$, and $d$. }
    (a) Two‐point correlation $C^z_d(t=T)$ at the final time $T=10$, plotted for spatial separations $d=1,\dots,N-1$ (curves range from deep blue at $d=1$ to deep red at $d=24$); labels indicate the shortest and longest distances. Inset showing the distance‐averaged peak correlation $\frac{1}{N-1}\sum_{d=1}^{N-1}\max_t\bigl|C^z_d(t)\bigr|$ versus $\gamma$. (b) Maximum correlation across distances as a function of $t$. The data ranges for $\gamma \in [0, 0.3]$ with step size of $0.05$.}
    \label{fig:correlations}
\end{figure}

We then proceed to the relation between the \textit{correlation length} and simulation complexity. Spin correlations serve as key indicators of the required computational resources as the presence of long-range correlations across distant sites generally means increased complexity \cite{prosen2008quantum}. Crucially, the correlation length bounds the number of measurements. The spin correlation function encodes these spatial dependencies
\begin{equation}
c_{ij} = \tr(\sigma^z_i \sigma^z_j \rho) - \tr(\sigma^z_i \rho) \tr(\sigma^z_j \rho), \nonumber
\end{equation}
where $i$ and $j$ denote different sites. Fig.~\ref{fig:correlations}(a) shows the correlation length $C^z_d(t)$ at distances $d \in {1,\dots, N-1}$ for a fixed long evolution time. The results demonstrate a systematic decrease in $C^z_d(t)$ as $\gamma$ increases. This behaviour indicates suppression of long-range correlations, and consequently, a reduction in the complexity of simulating strongly dissipative open quantum systems. Besides, to provide a broader perspective, Fig.~\ref{fig:correlations} shows the \textit{average} spatial correlations, which more clearly highlight the reduction of correlation length---and hence simulation complexity---with increasing dissipation at fixed time. Here we present results at relatively long times, where this behaviour is most pronounced, although a similar (albeit less drastic) trend is already visible at shorter times.
Since, according to section \ref{sec:complexity}, the running time for simulating open quantum system dynamics on a quantum computer is bounded by the {maximum correlation length} across all distances at each timestep, we also compute this quantity throughout the evolution of $S$ for each $\gamma$, see Fig. \ref{fig:correlations}(b). We note that correlations are not necessarily suppressed by long-time evolution alone: due to the oscillatory behaviour of the Hamiltonian, they can persist or even revive at later times. However, dissipation introduces a damping effect that becomes increasingly dominant with time, leading to a systematic reduction of correlations. This trend is consistent with the average correlation results in Fig. \ref{fig:correlations}(a), where dissipation clearly shortens the correlation length. Together with the analytical scaling in section~\ref{sec:complexity}, which links time complexity to the correlation length, these results indicate that dissipation effectively reduces the simulation complexity---even if correlations themselves do not always decay monotonically with total time evolution.

We also investigate time-dependent Lindbladian dynamics in Supplementary Section~\ref{sec:app_time_dep}, where either the dissipation strength, the Pauli operators, or the sites on which the jump operators act vary during evolution. Overall, we observe qualitatively similar behaviour to the results discussed here, with no significant deviations. At the same time, we find that randomised application of jump operators can accelerate relaxation processes, for instance by enhancing the decay of correlations, thereby modifying the system’s dynamical pathways. This further supports the generality of our conclusions and highlights the potential of such protocols for applications such as dissipative quantum state preparation, where time-dependent Lindbladians are often required to engineer the desired steady state~\cite{lin2025dissipative}. Finally, we wish to remark that while the discussion in this section has focused on linear Lindbladian dynamics, nonlinear master equations could be even more challenging to simulate classically, as their explicit matrix representation is already difficult to construct. Importantly, the quantum algorithm we introduce in this work can efficiently simulate both linear and nonlinear evolutions, making our conclusions directly relevant to the broader setting.

 \begin{figure}[t]
    \centering
    \includegraphics[width=1\linewidth]{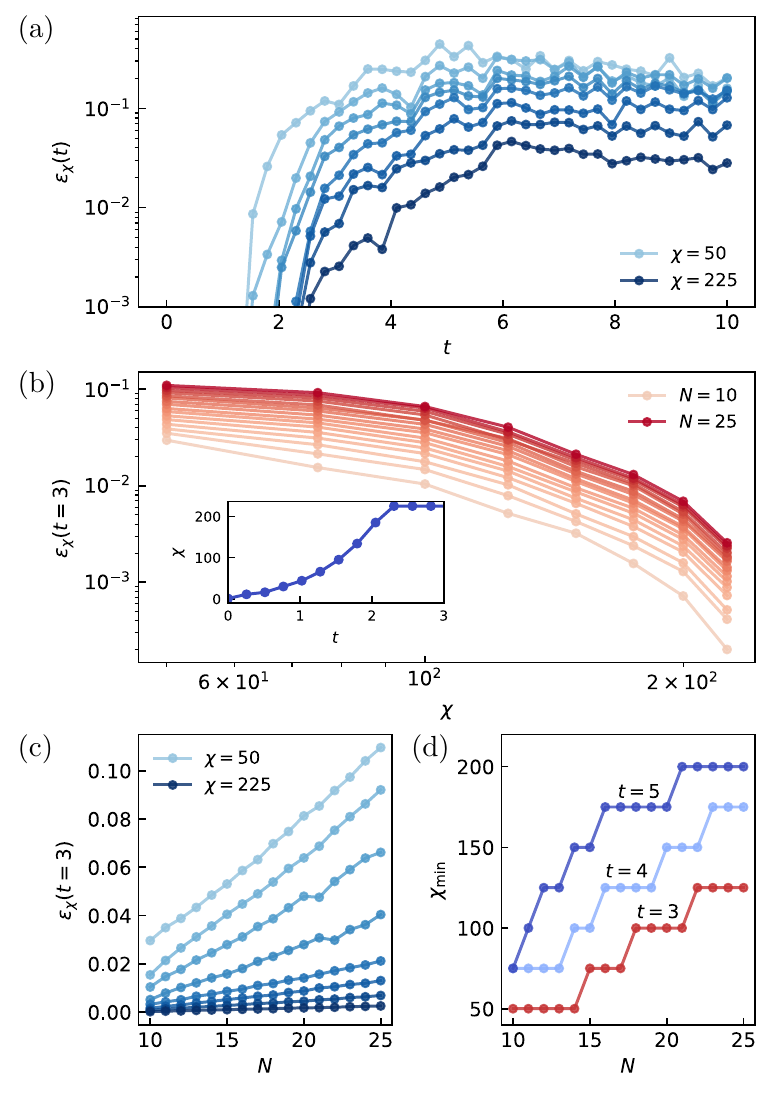}
    \caption{\textbf{Tensor network simulations for estimating the complexity of classical simulations.} In all panels $\gamma=0.05$. We compute the correlation error, indicative of the classical simulability of $S$: (a) as a function of $t$ for $N=25$ and $\chi\in[50,225]$, (b) as a function of $\chi$ for $N \in [10,25]$ and $t=3$; inset: growth of $\chi$ over $t$, and (c) as a function of $N\in[10,25]$ for $\chi\in [50,225]$ and $t=3$. (d) Minimum $\chi_{\min}$ to reach $\epsilon_{\chi_{\min}}\le 0.05$ as a function of $N \in [10,25]$ for $t \in \{3,4,5\}$.}
    \label{fig:bond_dimension}
\end{figure}

\noindent
\textbf{Classical runtime analysis.}
We now analyse the classical resource requirements---runtime and bond dimension---and provide simulation results to illustrate how simulation errors depend on time $t$, system size $N$, and bond dimension $\chi$. Generally, we expect simulation accuracy to improve for larger $\chi$ at the cost of increased computational runtime. The objective here is twofold: (i) to determine whether the growth trend differs between quantum and classical algorithms, and (ii) to assess whether classical algorithms exhibit rapid growth in complexity with increasing $N$ and $t$.

Figure \ref{fig:bond_dimension} illustrates the classical simulability of open quantum systems through that accuracy-complexity trade-off. For smaller systems, we quantify simulation errors by comparing results at finite $\chi$ against the ground-truth solution computed using a sufficiently large bond dimension, chosen here as $\chi=250$ to ensure convergence of the expectation values of the observable (see Fig.~\ref{fig:obs_bond_dimension} in Supplementary Section \ref{sec:app_tensor_network}). The accuracy of TEBD simulations at given $\chi$ with respect to the ground truth is measured via spin-spin correlation functions, with deviations quantified as the correlation error
\begin{equation}
    \varepsilon =\left( \frac{\sum_{ij}\left(c_{ij}-\tilde c_{ij}\right)^2}{\sum_{ij} \tilde c_{ij}^2}\right)^{1/2} ,
\end{equation}
where $\tilde c_{ij}$ are the ground-truth correlations. In Fig.~\ref{fig:bond_dimension}(a), we show how the correlation error $\varepsilon$ evolves over $t$ for $N=25$ and $\chi \in [50, 225]$. As expected, larger $\chi$ consistently yield better accuracy, especially at longer times. Since $S$ eventually reaches a steady state, $\varepsilon$ saturates to a constant value, which is why we think the error remains bounded over long simulation times. In a similar manner, in Fig.~\ref{fig:bond_dimension}(b) we fix $t = 3$ and examine $\varepsilon$ with respect to $N$ for $N \in [10,25]$. The results confirm the linear growth observed earlier. The main contribution to the complexity of the algorithm, however, arises from its scaling with $\chi$, as it determines the fidelity between our MPS state and the evolved quantum state. In Fig.~\ref{fig:bond_dimension}(c), $\varepsilon$ is plotted on a logarithmic scale as a function of $\chi$ for $N \in [10,25]$ and fixed $t=3$. Evidently, for a given $\chi$, smaller systems exhibit smaller $\varepsilon$, making them easier to simulate. For practical purposes, in Fig. \ref{fig:bond_dimension}(d) we set an error threshold $\epsilon_{\chi_{\min}}\le 0.05$ and identify the minimal bond dimension $\chi_{\text{min}}$ necessary to reach that threshold for every $N$. Naturally, $\chi_{\text{min}}$ depends on $t$, as errors consistently increase with longer simulation times, illustrated clearly in Fig. \ref{fig:bond_dimension}(d). Overall, as expected, we see that larger $\chi$ yield more precise results, progressively approaching exact solutions. This confirms that dynamics can be approximated with arbitrary accuracy given sufficiently large $\chi$. In our case, we find that $\chi = 200$ is sufficient to reach the error threshold for the system sizes considered, so working with samples up to $\chi=250$ is well justified. This is consistent with the results in Supplementary Section \ref{sec:app_tensor_network}, where we observe clear convergence of the observables with increasing bond dimension, further supporting that $\chi=250$ provides a reliable estimate.

\begin{figure}[!t]
    \centering
    \includegraphics[width=1.0\linewidth]{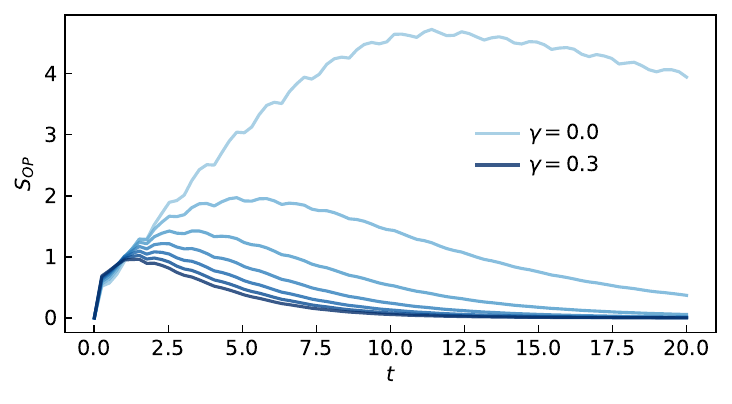}
    \caption{\textbf{Operator entanglement entropy growth under time evolution.} Operator entanglement entropy as a function of $t$ for $N=25$, $\chi = 250$, and $\gamma \in [0, 0.3]$ with step size of $0.05$. Each data point is obtained by calculating the operator space entanglement entropy of the bipartition between the two halves of the MPS.}
    \label{fig:entropy}
\end{figure}

The classical simulability of TNs is known to be fundamentally constrained by the entanglement entropy of the quantum state. To assess this limitation, we analyse the growth of entanglement entropy as an indicator of computational complexity.
In Fig.~\ref{fig:entropy}, we study the operator space entanglement entropy of the time-evolved state. We find that although the entanglement entropy initially increases, it then exhibits qualitatively different behaviour depending on $\gamma$. For $\gamma$ above a certain threshold, the entanglement entropy decreases over time, consistent with our expectations and the results presented by \cite{wellnitz2022rise}. However, when considering simulation complexity, a more relevant metric is the bond dimension $\chi$ required to represent the state. 
Surprisingly, we find that $\chi$ remains largely constant across different values of $\gamma$. This result is somewhat counter-intuitive compared to the behaviour of the entanglement entropy. The key difference lies in the fact that the entanglement entropy characterizes the final state, whereas simulating the full dynamics requires tracking the entire evolution, which can be more demanding. As a result, $\chi$ does not decrease with $t$ even for different values of $\gamma$.

To give a brief definition of the operator entanglement, we define our density matrix in MPO form as a Schmidt decomposition between the left $L$ and right $R$ block
\begin{equation}
    \hat \rho = \sum_i \lambda_i \hat o^{[L]}_i \hat o^{[R]}_i,
\end{equation}
where $\tr (\hat o^{[L/R]}_i \hat o^{[L/R]}_j)=\delta_{ij}$ and $\lambda_i$ denotes the Schmidt coefficients. Then, define the operator entanglement as 
$
    S_{\text{OP}} =-\sum_i \lambda_i^2 \log(\lambda_i^2).
 $ $S_{\text{OP}}$ is not strictly connected with quantum entanglement between blocks but it is an important quantity nonetheless that gives insights into the efficiency of the MPO representation. Recall that, for the purpose of our simulation, we have encoded the density matrix MPO into a MPS with double the quantum dimension. In this case, the operator entropy of $\rho$ is equivalent to calculating the entanglement entropy of the evolved MPS, carefully handling the extra physical dimension.

A noteworthy observations is that the complexity of TNs does not decrease with $t$ and $\gamma$ in certain regimes, in stark contrast to the complexity associated with quantum algorithms. This suggests there may exist a range of dissipation strength, where the separation between quantum and classical computational complexity is maximised.

\section{Discussion}
In this work, we examined the algorithmic complexity of simulating open quantum systems under the conjecture that, beyond a threshold timescale, Lindbladian dynamics may be easier to simulate than Hamiltonian dynamics. Instead of relying on dilation methods that obscure reduced-system complexity \cite{hu2020quantum,ticozzi2017quantum,schlimgen2022quantum}, we focused on correlation length and mixing time, employing a quantum jump approach---via QITE for the quantum simulations and TNs for the classical simulations. We introduced \sun{two post-selection-free algorithms that simulate both linear and nonlinear Lindbladians without ancilla overhead and normalisation issue~\cite{chan2023simulating,kamakari2022digital} of vectorisation by averaging the trajectories. The stochastic, continuous simulation method enables stochastic simulations of open dynamics, which may serve as a promising protocol for digital-analogue quantum simulations. Our framework extends naturally to non-Markovian dynamics defined on an extended Hilbert space.}

A key feature of open quantum systems is the competition between dissipative and unitary dynamics, where dissipation can either increase or suppress correlations or entanglement, thereby directly influencing the system's complexity~\cite{verstraete2009quantum}. The empirical results found so far suggest this behaviour is not universal, but in fact can highly depend on the type of state, the underlying dynamics, and the specific form of dissipation, as numerically found in \cite{prosen2008quantum,wellnitz2022rise}. 
To quantify this, we consider a normalised Lindbladian and track the correlation decay during the evolution. We analyse the algorithmic complexity in relation to the \textit{averaged} region- and time-dependent correlation length in Eq.~\ref{eq: averaged}, which is dynamic, and strictly smaller than the standard correlation length (e.g.,  \cite{brandao2015exponential,brandao2019finite}). From a computational perspective, our analytical result does not preclude the possibility that an intermediate time scale may exist, where dissipation increases the complexity beyond that of its closed counterpart.
In particular, in the main part of this work we showed numerically that stronger dissipation shortens mixing time and reduces correlation length. This result is similar to findings in the SYK model, where Krylov complexity was used to quantify the complexity growth, and dissipation was shown to suppress it compared with the closed counterpart \cite{bhattacharjee2023operator}. 

An noteworthy observation is that we observed a contrast between classical and quantum resource scaling: while TN simulations exhibit super-polynomial growth, quantum resources scale much favourably. Interestingly, dissipation suppresses correlations but does not reduce the TN bond dimension, showing there is a separation between quantum and classical computational complexity. This finding reveals the competition between dynamical correlations and transient entanglement and the distinction between them.

Our work provides insight into how dissipation shapes both classical and quantum algorithmic complexity via physically meaningful measures. These results may inform us in designing more efficient algorithms for simulating open quantum systems, and identifying regimes where quantum algorithms can outperform classical ones. Some interesting directions for future investigation emerge. Specifically, in preparing many-body ground states, which is known to be challenging even for quantum computers~\cite{lee2023evaluating},  a key question for dissipative quantum algorithms is how to reduce the complexity of open dynamics and the mixing time~\cite{lin2025dissipative,ding2024single}.  In Supplementary Section \ref{sec:app_time_dep}, we highlighted the role of randomness in the steady-state properties of local observables. Specifically, we demonstrated that applying jump operators randomly, rather than at fixed boundary sites, accelerates dissipative state preparation, suggesting that randomised dissipation protocols can be more efficient. This result indicates that one may exploit such dynamics to algorithmically design randomised dissipation protocols that systematically reduce mixing times, and thus enable efficient preparation of many-body ground states.
On a related note, it is also worthwhile to determine whether the observed polynomial scaling of quantum algorithms persists in realistic settings like those with nonlocal jump operators or correlated finite-temperature environments, which is an interesting research direction for ground-state preparation. 




\begin{acknowledgements}
We would like to thank Yuguo Liu, Yukai Guo and Weitang Li for the useful discussion related to open quantum dynamics simulation. 
 This research receives support from Schmidt Sciences, LLC. L.V.E. was supported by the Clarendon Fund. 
 We would also like to acknowledge funding from the UK EPSRC through EP/Z53318X/1. 
\end{acknowledgements}


\section*{Author contribution}
J.S. conceived the idea. L.V.E., A.Y., and J.S. developed the theoretical aspects of the project.  L.V.E. developed the source code and performed numerical simulations. J.S. carried out the theoretical analysis to support the study. L.V.E., A.Y. and J.S. wrote the manuscript.

\section*{Data availability}

All data needed to evaluate the conclusions in
the paper are present in the paper and the Supplementary Information. The source data is available by request from the authors.

\section*{Code availability}

The code is available by request from the authors.

\section*{Competing interests}

The authors have no competing interests.

\newpage
\clearpage

\appendix

\widetext
 
\section*{Supplementary Information}

\setcounter{section}{0}
\setcounter{figure}{0}
\setcounter{equation}{0}

\renewcommand{\thesection}{S\arabic{section}}
\renewcommand{\theequation}{S\arabic{equation}}
\renewcommand{\thefigure}{S\arabic{figure}}
\renewcommand{\thetable}{S\arabic{table}}

\renewcommand{\addcontentsline}[3]{\oldaddcontentsline{#1}{#2}{#3}}

\tableofcontents

\section{Formulation of open quantum dynamics}
 
In Supplementary Information, the hat notation is used to denote the terms in the extended space and the operators as well.

\subsection{Vectorisation of the reduced density matrix under linear Lindblad evolution}
\label{sec:app_vector_density}

To simulate the quantum dynamics governed by a linear Lindbladian, we encode the density matrix as a pure state through vectorisation. This transformation allows us to represent mixed-state evolution in a form suitable for quantum circuits. The same technique is also applied for classical TN simulations of open quantum systems, as described in the main text
The mapping sends $\rho$ to a vector in the Liouville space
\begin{equation}
|\rho\rangle\rangle = \sum_{i,j} \rho_{ij} |i\rangle \otimes |j\rangle.\nonumber
\end{equation}
Under this transformation, the Lindblad master equation takes a Schrödinger-like form
\begin{equation}
\frac{d}{dt}|\rho\rangle\rangle = \hat{\mathcal{L}} |\rho\rangle\rangle,\nonumber
\end{equation}
where the Liouvillian superoperator $\hat{\mathcal{L}}$ acts linearly as
\begin{equation}
\begin{aligned}
\hat{\mathcal{L}} =& -i(\mathbb{I} \otimes H - H^{\mathsf T} \otimes \mathbb{I}) \\
&+ \sum_k \left( L_k \otimes L_k - \tfrac{1}{2} \mathbb{I} \otimes L_k^\dagger L_k - \tfrac{1}{2} L_k^\dagger L_k \otimes \mathbb{I} \right).
\end{aligned}
\end{equation}
To compare open-system ($\gamma > 0$) with unitary ($\gamma = 0$) dynamics on equal footing, we consider a normalised Liouvillian
\begin{equation}
\hat{\mathcal{L}} = -i ( \lambda_1 H_1 - i \lambda_2 H_2),
\end{equation}
where $||H_1|| = ||H_2|| = 1$ and $\lambda_1^2+\lambda_2^2 = 1$, with $H_1$ and $H_2$ is the matrix representation of the superoperator.
In analogy with the operator $H_{\text{eff}}$ in Eq.~\ref{eq:Heff}, the time evolution of the vectorised density matrix can be written as
\begin{equation}
\ket{\rho(t)}\rangle = \exp[-i(  \lambda_1 H_1 t - i  \lambda_2 H_2 t)] \ket{\rho(0)}\rangle.
\end{equation}
The derivation then proceeds by applying a first-order Trotter decomposition, as an example
\begin{equation}
\begin{aligned}
\ket{\rho(t)}\rangle =&\,
   \Big[e^{-i \lambda_1 H_1 \delta t}\,
        e^{- \lambda_2 H_2 \delta t}\Big]^{T / \delta t} \ket{\rho(0)}\rangle
   + \mathcal{O}\!\left( T \delta t \right) ,\nonumber
\end{aligned}
\end{equation}
which adopts the same form as in the quantum jump situation.
In Supplementary Section~\ref{app:totalerror}, $K$th-order Trotter decomposition is applied.
The dynamics of the vectorised state is then approximated as alternating applications of unitary operations and operations derived from the QITE approach.

\subsection{Pseudomode formulation of non-Markovian dynamics and beyond}
\label{sec:appA}

In open quantum dynamics, when the environment has a non-negligible memory, we need to go beyond the Lindblad master equation to describe the reduced system dynamics.
A numerically exact way to deal with the non-Markovian dynamics is to use the HEOM~\cite{tanimura2020numerically}, which contains a hierarchy that connects to different auxiliary density operators. Similarly, Yan et al. \cite{yan2016dissipation,yan2014theory} proposed the dissipaton equation of motion, where quasi-bosonic particles (i.e., bosonic dissipatons) are introduced as auxiliary density operators to encode the non-negligible system-environment interactions. Later, this was extended to the DQME formalism, where instead of a hierarchical structure, the reduced system dynamics is incorporated into a single dynamic equation, and then extended to a second-quantised form of DQME, where the reduced density tensor of the dissipation embedded system $\tilde{\rho} \in \mathcal{H}_S \otimes \mathcal{H}_S ^* \otimes \mathcal{H}_F$ (with the Fock space $ \mathcal{H}_F$) can be implemented by using linear combination of unitaries algorithms.

As we discussed in the main text, one way of unravelling the influence functional of the system-bath interaction is to decompose the bath correlations into a finite set of modes, cf. Eq. \ref{eq:c_env}. This also aligns with the idea that we can rewrite the memory kernel of the reduced
dynamics into a set of time-local interactions between the system and a finite number of pseudo-modes, and use the pseudo-mode Lindbladian~\cite{tamascelli2018nonperturbative} as an approximation to the non-Markovian dynamics when $d_k \in \mathbb{R}^1$. 

Along those lines, a few approaches have been proposed for simulating non-Markovian dynamics based on this extended space, see \cite{li2024toward}. From now on, we mainly follow the method in~\cite{xu2023universal} which unravels influence functions using complex-valued auxiliary paths. We will thus provide a self-contained procedure for representing non-Markovian dynamics using the Lindblad-type equation in an extended space.

The reduced density operator ${\rho}_S$ is expressed as a functional integral over paths supported by a Keldysh contour
\begin{equation}
\rho_{S}^{ \pm}(t)=\int \mathcal{D}\left[q_{S}^{+}, q_{S}^{-}\right] e^{iS\left[q_{S}^{+}, q_{S}^{-}\right]}e^{F\left[q_{S}^{+}, q_{S}^{-}\right]} \rho_{S}^{ \pm}(0),
\end{equation}
and the bare action factor  $e^{iS\left[q_{S}^{+}, q_{S}^{-}\right]}$ captures the quantum dynamics in the absence of a bath. Here, $S[q^+_S]$ and $S[q^-_S]$ are the corresponding actions associated with, respectively, forward and backward paths $q^{\pm}_S(t)$. Formally 
\begin{equation}
\begin{aligned}
F\big[ q_{S}^{+}, q_{S}^{-}\big]
&= -\int_{0}^{t} ds\int_{0}^{s}d\tau
\big( q_{S}^{+}(s)- q_{S}^{-}(s)\big)\\
& \qquad\times\big(C^{\mathrm{env}}(s-\tau)  q_{S}^{+}(\tau)-C^{\mathrm{env} *}(s-\tau) q_{S}^{-}(\tau)\big).
\nonumber
\end{aligned}
\end{equation}
In \autoref{sec:methods} in the main text, we presented a common approach in this regime, where we represented the environment through its bath correlation function Eq. \ref{eq:c_env}, and expanded it as a finite sum of complex exponentials with real amplitudes $d_k$ and complex frequencies $z_k$. This decomposition enabled the mapping of the system’s dynamics onto an extended space $\Gamma$. Building on that, we have
$$
\begin{aligned}
\rho_{S}^{ \pm}(t)=\int & \mathcal{D}\left[q_S^{+}, q_S^{-}\right] \prod_{k=1}^K \mathcal{D}\left[\phi_k^*, \phi_k ; \psi_k^*, \psi_k\right] e^{iS\left[q_{S}^{+}, q_{S}^{-}\right]} \\
& \times e^{i S_k\left[\phi_k, \phi_k^*, q_S^{+}, q_S^{-}\right]+i \bar{S}_k\left[\psi_k, \psi_k^*, q_S^{+}, q_S^{-}\right]} \rho_{S}^{ \pm}(0),
\end{aligned}.
$$
As derived in \cite{xu2023universal} the actions are
\begin{equation}
\begin{aligned}
S_k\left[\phi_k, \phi_k^*, q_{S}^{+}, q_{S}^{-}\right] & =\int_0^t d\tau\left(i \phi_k^* \partial_\tau \phi_k-H_k\left(\phi_k, \phi_k^*\right)\right), \\
\bar{S}_k\left[\psi_k, \psi_k^*, q_{S}^{+}, q_{S}^{-}\right] & =\int_0^t d\tau\left(i \psi_k^* \partial_\tau \psi_k-\bar{H}_k\left(\psi_k, \psi_k^*\right)\right),
\end{aligned}
\end{equation}
and
$
H_k\left(\phi_k, \phi_k^*\right)=-i z_k \phi_k^* \phi_k-i \sqrt{d_k}\left(q_{S}^{+}-q_{S}^{-}\right) \phi_k+i \sqrt{d_k} q_{S}^{+} \phi_k^*,
$
and the intermediate interaction terms are introduced
$
\bar{H}_k\left(\psi_k, \psi_k^*\right)=-i z_k^* \psi_k^* \psi_k-i \sqrt{d_k^*}\left(q_{S}^{-}-q_{S}^{+}\right) \psi_k+i \sqrt{d_k^*} q_{S}^{-} \psi_k^*.
$

\sun{
The idea behind dissipation-embedded quantum master equation~\cite{li2024toward} or QD-MESS~\cite{xu2023universal} is to introduce an extended space $\Gamma$ and derive the dynamics of states in the extended space. Below, we provide a few details in arriving at the evolution of states given by Eq. \ref{eq:lindd_pm}.}
 
For a bosonic mode with annihilation and creation operators
\(\hat a,\,\hat a^\dagger\) and a normal-ordered Hamiltonian
\(\widehat H := H_N(\hat a^\dagger,\hat a)\),
the coherent-state matrix element of the short-time propagator is
\[
K(\phi_{n+1}^*,\phi_n;\Delta t)
   = \langle \phi_{n+1}|e^{-i\Delta t\,\widehat H}|\phi_n\rangle.
\]
Since \(\hat a|\phi_n\rangle = \phi_n|\phi_n\rangle\) and
\(\langle\phi_{n+1}|\hat a^\dagger=\langle\phi_{n+1}|\phi_{n+1}^*\),
expanding to first order in \(\Delta t\) gives
\[
\begin{aligned}
\langle\phi_{n+1}|e^{-i\Delta t\widehat H}|\phi_n\rangle
 &\approx
 \langle\phi_{n+1}|\phi_n\rangle
   - i\Delta t\,\langle\phi_{n+1}|\widehat H|\phi_n\rangle\\[3pt]
 &= e^{\phi_{n+1}^*\phi_n}
    \left[1 - i\Delta t\,H_N(\phi_{n+1}^*,\phi_n)\right] ,
\end{aligned}
\]
and we have the standard short-time kernel identity
\begin{equation}
\langle\phi_{n+1}|e^{-i\Delta t\widehat H}|\phi_n\rangle
   = \exp\!\left(\phi_{n+1}^*\phi_n - i\Delta t\,H_N(\phi_{n+1}^*,\phi_n)\right)
\label{eq:shorttime}
\end{equation}
up to second order. Now let's define
\[
W(t)
   = \mathcal T
     \exp\!\left(-i\!\int_0^t \widehat H_{\mathrm{tot}}(\tau)\,d\tau\right)
     W(0) ,
\]
which is the propagator acting on the extended density-like object $W(0)$, but compared to the Schrödinger equation, it includes auxiliary-mode degrees of freedom.
Differentiating with respect to time yields
\begin{equation}
{\dot W(t) = -i\,\widehat H_{\mathrm{tot}}(t)\,W(t)},
\label{eq:Wdot}
\end{equation}
which is the operator form of the evolution of the extended object \(W(t)\) in Liouville-Fock space. Consider now one auxiliary mode with complex frequency
\(z_k  \).
The quadratic (free) part of the coherent-state action for this mode is
\[
S_k^{\mathrm{free}}[\phi_k,\phi_k^*]
   = \int_0^t d\tau\,
     \Big(i\phi_k^*\dot\phi_k - H_{N,k}^{\mathrm{free}}(\phi_k^*,\phi_k)\Big),
\]
with the normal symbol
\(
H_{N,k}^{\mathrm{free}} = -i z_k \phi_k^*\phi_k.
\)
Hence, the Lagrangian corresponding to this free part is
\begin{equation}
L_k^{\mathrm{free}}
   = i\phi_k^*\dot\phi_k + i z_k \phi_k^*\phi_k ,
\label{eq:L0}
\end{equation}
where the term \(i\phi_k^*\dot\phi_k\) arises from the coherent-state overlap
between neighbouring time slices (see Eq. \ref{eq:shorttime}),
while the term \(i z_k \phi_k^*\phi_k\) comes from
  \(H_{N,k}^{\mathrm{free}}=-i z_k \phi_k^*\phi_k\). Replacing the c-number fields by their corresponding ladder operators,
\(\phi_k \mapsto \hat a_k,\ \phi_k^*\mapsto\hat a_k^\dagger\),
gives the normal-ordered Hamiltonian operator
\[
\widehat H_k^{\mathrm{free}} = -i z_k\,\hat a_k^\dagger \hat a_k.
\]
According to Eq. \ref{eq:Wdot}, the contribution of this mode
to the time derivative of \(W\) is
\[
-i\widehat H_k^{\mathrm{free}}W
   = -i(-i z_k)\,\hat a_k^\dagger \hat a_k W
   = -z_k\,\hat a_k^\dagger \hat a_k W .
\]
Similarly, for the backward branch, one obtains
\(-z_k^*\,\hat b_k^\dagger \hat b_k W\). Therefore, we have the generator $\hat{\Gamma}_k = - z_k \hat{a}_k^{\dagger} \hat{a}_k - z_k^* \hat{b}_k^{\dagger} \hat{b}_k$ that is introduced in the main text. 
Now the forward coherent fields correspond to operators acting from the left on the extended object. Specifically, the mapping to operators acting on \(W\) is
$
\phi_k \mapsto \hat a_k ,  
\phi_k^* \mapsto \hat a_k^\dagger \quad(\text{left action}),$  $
\psi_k \mapsto \hat b_k, 
\psi_k^* \mapsto \hat b_k^\dagger \quad(\text{right action}).
 $
Forward system coordinate \(q_S^+\) maps to left multiplication by \(\hat q_S\), while backward coordinate \(q_S^-\) maps to right multiplication by \(\hat q_S\), which is
\[
q_S^+ \longmapsto \hat q_S\,W, \qquad q_S^- \longmapsto W\,\hat q_S.
\]
It is easy to check that $\left(q_{S}^{+}-q_{S}^{-}\right) \phi_k = [ \hat q_S, \hat{a}_k {W}]$. The dynamics of an extend state ${W}$ then reads 
\begin{equation}
\begin{aligned}
\dot{{W}}= & -i \mathcal{L}_{S} {W}+\sum_k\left\{\hat{\Gamma}_k {W}+\sqrt{d_k} \hat{q}_{S} \hat{a}_k^{\dagger} {W}\right. \\
& \left.+\sqrt{d_k^*} \hat{b}_k^{\dagger} {W} \hat{q}_{S}-\left[\hat{q}_{S}\left(\sqrt{d_k} \hat{a}_k-\sqrt{d_k^*} \hat{b}_k\right) {W}\right]\right\}.
\end{aligned}
\end{equation}
When considering the decomposition 
\begin{equation}
{W}(t)=\sum_{\mathbf{m}, \mathbf{n}} {\rho}_{\mathbf{m}, \mathbf{n}}(t)|\boldsymbol{m}, \boldsymbol{n}\rangle, \nonumber
\end{equation} 
we will arrive at the HEOM of ${\rho}_{\mathbf{m}, \mathbf{n}}(t)$, which could be shown to be equivalent to that in~\cite{xu2022taming}. One may then apply the similarity transformation $\mathcal{S}$ discussed in the main text that mixes $\hat{a}_k$ and $\hat{b}_k$, allowing the dynamics to be expressed in Lindblad form. Besides, recall that to enable implementation on a quantum computer, the state was further defined using only half of the auxiliary bosonic modes in a dual Fock space ${\rho} = \sum_{\mathbf{m},\mathbf{n}} \ket{\mathbf{m}}\rho_{\mathbf{m},\mathbf{n}} \bra{\mathbf{n}}$. In doing so, the dynamics is mapped to a quasi-Lindblad
\begin{equation}
\begin{aligned}
\label{eq:NM_general_DM}
\partial_t{{\rho}}_{\mathcal{S}}=&-i\left[\hat{H}_{\mathrm{pm}}, {\rho}_{\mathcal{S}}\right] + \mathcal{D_{\rm pm}}[{\rho}_{\mathcal{S}}]\\
&+ \sum_{k=1}^K \frac{\IM(d_k)}{\RE(d_k)}\left([\hat{a}_k, {\rho}_{\mathcal{S}}] \hat{q}_{S}-\hat{q}_{S}[\hat{a}_k^{\dagger}, {\rho}_{\mathcal{S}}]\right).
\end{aligned}
\end{equation}

One can find that the density matrix is not positive but is Hermitian and trace-preserving $\tr({\rho}_{\mathcal{S}}) = 1$. It is easy to check that the reduced density is also trace-preserved $\tr({\rho}_{S}) = 1$. When $d_k$ is constrained to be real, then we recover the Lindbladian \ref{eq:lindd_pm} in the main text. It is moreover interesting to compare this with the one obtained from pseudomode theory in \cite{tamascelli2018nonperturbative,nusseler2022fingerprint,menczel2024non}, where $d_k$ is real. The dynamics of the state of $S$ is given by $$\rho_S(t) = \sum_{ \mathbf{n}} \braket{\mathbf{n}|\rho_{\mathcal{S}}(t)| \mathbf{n}}  = \tr_{\backslash S} \rho_{\mathcal{S}}(t),$$ 
where it directly encodes the real amplitude $d_k$ and  the complex frequency $z_k$ into the Lindbladian. In the more general case with $\IM(d_k) \neq 0$, we can still simulate the dynamics with a vectorised form. Specifically, c.f. Eq.~(37) in \cite{xu2023universal}
\begin{equation}
    \begin{aligned}
\frac{d}{dt}\,|\rho_S\rangle\rangle
=& \hat{\mathcal{L}}_S  |\rho_S\rangle\rangle = 
\Big[
 -i\big(\mathbb{I} \otimes \hat H_{\mathrm{pm}}
   - \hat H_{\mathrm{pm}}^{\mathsf T} \otimes \mathbb{I}\big) \\
   &
 + \sum_k \mathcal{L}_{C,k}
 + \sum_k \mathcal{L}_{D,k}
\Big]\,|\rho_S\rangle\rangle,
\label{eq:NM_vector}
\end{aligned}
\end{equation}
where the individual superoperators are given as
\begin{equation}
\begin{aligned}
\mathcal{L}_{D,k}
&= \RE(z_k)
\Big(
 \hat 2 a_k^* \otimes \hat a_k
 -   \mathbb{I} \otimes (\hat a_k^\dagger \hat a_k)
 -  (\hat a_k^\dagger \hat a_k)^{\mathsf T} \otimes \mathbb{I}
\Big),
\end{aligned}
 \end{equation}
 and
 \begin{equation}
\begin{aligned}
\mathcal{L}_{C,k}
=& \frac{ \IM(d_k) }{\sqrt{ \RE(d_k)} }
\Big[
 \hat q_S^{\mathsf T} \otimes \hat a_k
 - (\hat q_S \hat a_k)^{\mathsf T} \otimes \mathbb{I}
 \\&- \mathbb{I} \otimes (\hat q_S \hat a_k^\dagger)
 + (\hat a_k^\dagger)^{\mathsf T} \otimes \hat q_S
\Big].
\end{aligned}
 \end{equation}
In this form, we can still express $\hat{\mathcal{L}}_S$ as a normalised Liouvillian
\begin{equation}
\hat{\mathcal{L}}_S = -i (   H_1 - i   H_2),
\end{equation}
where $H_1 = \mathbb{I} \otimes \hat H_{\mathrm{eff}}
   - \hat H_{\mathrm{eff}}^{\mathsf T} \otimes \mathbb{I}$ and $H_2 = \sum_k \mathcal{L}_{C,k} + \mathcal{L}_{D,k}$ is the matrix representation of the superoperators. Note that this is different from the DQME-SQ in \cite{li2024toward}, where the generator is not Hermitian but trace preserving. It may be worth noting that one may adopt other unravelling methods to get the effective Hamiltonian, e.g.,~\cite{ke2022hierarchical}.



\section{Error bound of the evolution and the algorithmic complexity}
\label{app:totalerror}

Below we analyse the simulation complexity of open quantum system dynamics.
\sun{The simulation is composed of the dynamics governed by $H_{\rm eff}$ and the quanutm jump. The dynamics under $H_{\rm eff}$ is composed of both real-time evolution and imaginary-time evolution. As we shall see, the realisation of the quantum jump also uses imaginary-time evolution. }

We first discuss the complexity in simulating the dynamics under $H_{\rm eff}$. 
We divide the total time evolution $T$ into $N_2$ segments of duration $\delta {t} = T/N_2$. For the non-Hermitian component $H_2$, non-commuting terms introduce a Trotter error in QITE. To control Trotter error, we further divide each $\delta {t}$ into $N_1$ sub-steps. 
Here, we consider a $K$th-order Trottersation at all levels (with even $K$).
The error sources include Trotter error in dividing the evolution into Hermitian and non-Hermitian parts, Trotter error in real-time evolution, and for unitary decomposition of the imaginary-time evolution operator, each with its own error set by $\epsilon/3$.
The number of sub-steps $N_1$ is set by the condition
\begin{equation}
\frac{N_2 (\lambda_2 \delta {t})^{K+1}}{N_1^K} = \frac{\varepsilon}{3} = \mathcal{O } (\varepsilon),\nonumber
\end{equation}
which gives
\begin{equation}
N_1 = \mathcal{O} \left(\frac{(\lambda_2 T)^{ \frac{K+1}{K}}}{N_2 \varepsilon^{\frac{1}{K}}}\right).\nonumber
\end{equation}
The simulation error is governed by the average correlation length $\overline{C}$. For the effective non-Hermitian Hamiltonian $H_{\rm eff}$, we write
\begin{equation}
H_2 = \sum_{i=1}^{M_k} h_{2,i}.\nonumber
\end{equation}
Following the argument in \cite{motta2020determining}, let $\Psi_p$ denote the pure state at the $p$th step
\begin{equation}
\ket{\Psi_p} = \frac{ \prod_{k=1}^{M_k} e^{-\delta {t} h_{2,k}} \ket{\Psi_{p-1}}}{||\prod_{k=1}^{M_k} e^{-\delta {t} h_{2,k}} \ket{\Psi_{p-1}}||},
\end{equation}
with density matrix $\rho_p = \ket{\Psi_p}\bra{\Psi_p}$.
Consider a region $R_v$ consisting of all sites within distance $v$ of the support of $h_k$. The distance between successive reduced states satisfies
\sun{
\begin{equation} 
\begin{aligned} 
& \| \tr_{\setminus R_{v} } \rho_p - \tr_{\setminus R_{v} } \rho_{p-1}\|_1 \\ 
& \leq \sum_{i = 1}^{M_k} \|  \rho_{p, i+1}  - \rho_{p, i} \|_1 \\ 
&\leq \sum_{i = 1}^{M_k} e^{- \frac{v}{C_i}} \leq {M_k} e^{- \frac{v}{\bar{C}}},
\end{aligned} 
\end{equation}
where $ \rho_{p, i} := c_{p,i}^{-1} \prod_{k=i}^{M_k} e^{-2 \delta {t} h_{2, k}} \rho_{p-1}$, and $c_{p,i}$ denotes the normalisation factor. In the last line, we have used the result,  Lemma 9 of  \cite{brandao2015exponential}, which is used as a key lemma for the result in \cite{motta2020determining}. The last inequality relates the region-dependent $C_i$ to the average correlation length.
To be specific, we define $C_i$ as the minimal length such that Eq. \ref{eq: time_dep_corr} holds given all the possible pairs $(A_i, B_i) $, where $A_i$ is the region on which $h_{2,k}$ acts and $B_i$ is the region that has distance $v$ to $A_i$.
Therefore, the relevant parameter here is the average correlation length $\overline{C}$, rather than the maximal one.   
Choosing the domain size to be
$
v = \mathcal{O} \left( \bar{C} \ln\left( {M_k} N_{2} \varepsilon^{-1} \right) \right),
$
we can bound the error from approximating the Lindbladian by a sequence of unitaries.
Note the reason why we pair the terms in the second line is that we hope to have a large domain $v$  such that each individual term can be bounded by $e^{-v/C_i}$.}

The sample complexity per timestep for the non-Hermitian evolution is
\begin{equation*} 
\begin{aligned} 
S_q^{\rm NH}(\delta {t}) &= N_1 M_{K}  3^{  \max v ^{\mathrm{dim}} } \\ 
&= \mathcal{O} (N_1 N_2 M_K^2 e^{ \bar{C}^{\mathrm{dim}} } \varepsilon^{-1} ).
\end{aligned} 
\end{equation*}
At a higher level, $N_2$ is fixed by the Trotter error between $H_1$ and $H_2$
\begin{equation}
\epsilon_{\rm Trotter} = \mathcal{O } \left(  (\lambda_1 \lambda_2)^{K+1}   \frac{T^{K+1}}{N_{2}^K} \right) = \frac{\varepsilon}{3},
\end{equation}
so that
\begin{equation}
N_1 = \mathcal{O} ( \lambda_1^{-(1+\frac{1}{K})} ),
\qquad
N_2 = \mathcal{O}\left( \frac{(\lambda_1 \lambda_2 T)^{1+\frac{1}{K}}}{\varepsilon^{\frac{1}{K}}} \right).\nonumber
\end{equation}
Thus, the sampling cost per time step due to the non-Hermitian part is
\begin{equation*} 
S_q^{\rm NH}(\delta t) = \mathcal{O} \left( M_K^2  (\lambda_2 T )^{1+\frac{1}{K}} e^{  \bar{C}^{\mathrm{dim}} }  \varepsilon^{-(1+\frac{1}{K})}\right). 
\end{equation*}

In many physically relevant cases, environmental interactions are described by local Lindbladians with commuting terms in $H_2$. Then, no $N_1$ subdivision is required. With some derivation, the total complexity is given by
\begin{equation*} 
S_q^{\rm NH}(T) = \mathcal{O} \left( M_K^2  (\lambda_1\lambda_2 T )^{2+\frac{2}{K}} e^{  \bar{C}^{\mathrm{dim}} }  \varepsilon^{-(1+\frac{2}{K})} \right). 
\end{equation*}
Finally, note that $\overline{C} \leq \xi$, where $\xi$ is the universal correlation length of the system. Physically, $\overline{C}$ measures correlations at a fixed distance $d$ with all possible operators, whereas $\xi$ captures correlations at all distances. Our formulation sharpens the dependence on correlation length, since $\overline{C}$ can be much smaller than $\max_i C_i$. We note that the above expression characterises the total sample complexity accumulated over all steps. It also reflects the computational complexity associated with the stepwise growth of the circuit (with each block of duration $\delta t$).

The realisation of the quantum jump $L_k = U_k D_k V_k$, in particular $D_k \approx \exp( -H^{D_k} T^{D_k})$, is similar.
To approximate the diagonal operator $D_k$, we can choose to set the exponent $-H^{D_k} T^{D_k} = \sum_{a_{k, j} \neq 0 } \log(a_{k, j} ) \ket{j} \bra{j} - b_{D_k}  \sum_{a_{k, j} = 0 } \ket{j} \bra{j} $. Here, $b_k$ is an appropriately chosen constant. In case $\braket{D_k^2}$ is exponentially small, then one can still choose $b_k$ and output a zero vector (which is recorded for classical computation).
Otherwise, when $\braket{D_k^2} = \braket{\Psi_S | D_k^2| \Psi_S}$ is lower bounded by $b_k$, then one can choose to set that constant as
\begin{equation}
    b_k = \log(\varepsilon^{-1}) + \log(\braket{D_k^2}^{-1}) ,
\end{equation}
such that $\| D_k -\exp( -H^{D_k} T^{D_k} ) \|\leq \varepsilon $. Here, $\braket{D_k^2}$ can be directly measured at each timestep, and $H_{D_k}$ is local. For example, for $L_k = \sigma^+_k = \ket{1}\bra{0}$, it was shown in \cite{endo2020variational} that the effective Hamiltonian can be set as $H_{D_k} = (Z_k + \mathbb{I}_k)/2$, while $U_k = \mathbb{I}$ and $V_k = Z$. With this, one can use the QITE algorithm to find a unitary that approximates $D_k$.



\section{Space-time complexity bounds for simulating Lindbladian dynamics}
\label{app:spacetime}

To simulate an open quantum system with correlation length $\xi$ over time $T$, we may simply add more ancilla to mediate the relevant memory effects. The space-time complexity---defined as the total number of ancilla qubits multiplied by circuit depth---is found to be polynomial in $\xi$ and $T$. Here, we provide more details on this.

Assume that the Lindbladian dynamics arises from a unitary evolution of a joint system-bath state followed by tracing out the bath. That is
\begin{equation}
\frac{d\rho_S}{dt} = \mathrm{Tr}_B \left( -i [H_{SB}, \rho_{SB}(t)] \right),
\end{equation}
with system-bath interactions of the form $H_{SB} = \sum_j h_{j, B_j}$, where $h_{j, B_j}$ acts locally on site $j$ and its associated bath region $B_j$. Assume the bath is gapped or thermal, and that system-bath correlations decay exponentially in both space and time
\begin{equation}
|| \langle B_i(t), B_j(0) \rangle - \langle B_i(t) \rangle \langle B_j(0) \rangle| | \leq C e^{-|i-j|/\xi} e^{-t/\tau_c},
\end{equation}
where $\xi$ is the correlation length and $\tau_c$ is the bath memory (correlation) time. Discretise the evolution into $T$ timesteps of duration $\delta t$, so that $t = T \delta t$. At each timestep, the system evolves under a quantum channel $ 
\rho{(t+\delta t)} = \mathcal{E}_{\delta t}(\rho{(t)}).
$
Due to the finite memory time $\tau_c$, the bath correlations are negligible beyond a temporal window of $\tau_c$, implying that each $\mathcal{E}_{\delta t}$ can be modelled with ancillas that only retain correlations within the last $\tau_c / \delta t$ steps.
From the Stinespring dilation theorem, such a channel can be realised by a unitary interaction between the system and an ancilla of finite dimension, followed by a partial trace
\begin{equation}
\mathcal{E}_t(\rho) = \mathrm{Tr}_E \left[ U_t (\rho \otimes \ket{0}\bra{0}_E) U_t^\dagger \right].
\end{equation}
Finite bath \textit{memory time} $\tau_c$ means the environment (bath) only retains information about the $S$'s past for up to time $\tau_c$.
Any influence of $S$'s state before 
$t - \tau_c$ on its current evolution is exponentially suppressed. In other words, the bath acts approximately Markovian on timescales larger than $\tau_c$.

If the dynamics of $S$ were strictly Markovian, we would only need a single ancilla per timestep (and could even reuse it). However, in general cases, where $S$ exhibits non-Markovianity with a memory time $\tau_c$, we must retain information for up to $m = \tau_c / \delta t$ previous steps.
If  the bath memory lasts for a finite time 
$\tau_c = \mathcal{O}(1)$, the number of previous timesteps that affect the current step is bounded by 
\begin{equation}
    m = \tau_c/\delta t =  \tau_c N_{\text{steps}}/ T= \mathcal{O} ( ( \lambda T)^{ 1/K} \varepsilon^{-1/K}) , 
\end{equation}
where $\lambda$ denotes the interaction strength. 
This is the memory length for each timestep.
The total ancilla usage across all 
$N_{\text{steps}}$ timesteps is
$$N_{\text{steps}} * m =  \mathcal{O} ( (\lambda T)^{ 1+ 1/K} \varepsilon^{-1/K} ).$$
Note that in simulating an entire system of 
$N$ spatial sites, each site evolves under open-system dynamics, meaning it may couple to its own environment (e.g., local bath modes). 
In the worst case, each site needs its own set of ancilla qubits (or ancilla modes) to mediate interaction with the environment.
So the ancilla resource scales linearly with $N$ in the worst case. Due to locality and finite correlation length, each system site is influenced primarily by bath modes within a spatial region of radius $\xi$. Therefore, for each site at each timestep, we only need to keep track of a region of environment memory of size. Putting all this together, the total complexity (i.e., ancilla $\times$ circuit depth) is
$$ N * N_{\text{steps}} * m * \xi =   \mathcal{O} ( N \xi (\lambda T)^{ 1+ 1/K} \varepsilon^{-1/K}),$$ which is polynomial in correlation length and simulation duration. This validates the claim that, under assumptions of system-bath correlation decay, the cost of simulating open quantum systems with finite correlation length is efficiently bounded.

\section{Time-dependent Lindbladian}
\label{sec:app_time_dep}
Here, we provide details and illustrative results for two alternative dissipative protocols, where the Lindbladian dynamics is time-dependent in some way, focusing on the effects on both local magnetisation and correlations.

In the first protocol, the system evolves under a time-dependent Lindblad equation, where the dissipation strength $\gamma(t)$ varies smoothly during the evolution according to
\begin{equation}
    \gamma(t) = \gamma_\mathrm{max} \left| \sin\left( \frac{\pi t}{t_\mathrm{total}} \right) \right|.
\end{equation}
Here, $\gamma_\mathrm{max}$ sets the maximum dissipation rate reached during the dynamics. The Lindblad operators are fixed as $\sigma^\pm$ acting at the two ends of the chain, as in the main text.

\begin{figure}[t]
    \centering 
    \includegraphics[width=0.75\linewidth]{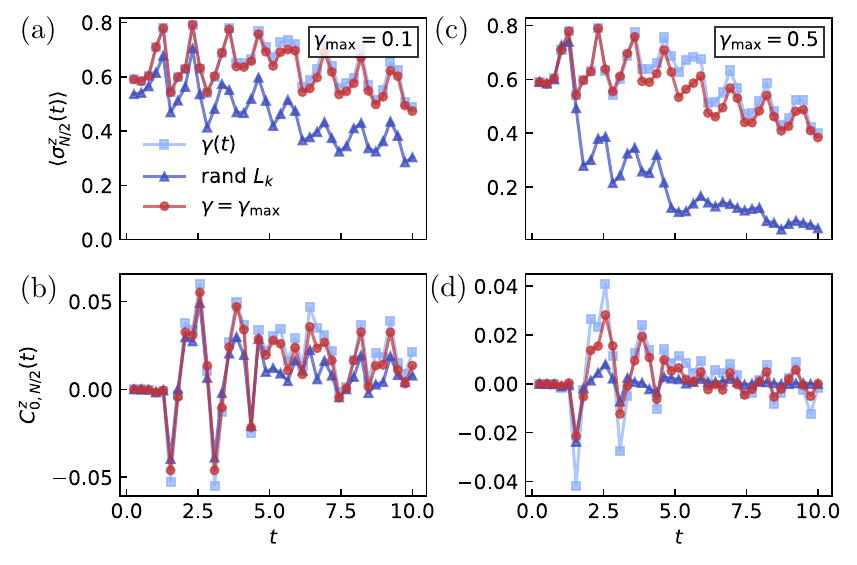}
    \caption{Dynamics under time-dependent and random-site Lindbladian protocols for system $S$ of size $N=11$ and maximum bond dimension $\chi=150$.
    Top row: Time evolution of the local magnetisation $\langle \sigma^z_{N/2}(t)\rangle$ at the center of the chain for (a) $\gamma_{\max}=0.1$ and (b) $\gamma_{\max}=0.5$.
    Bottom row: Connected correlation $C^z_{0,N/2}(t)$ between the ends and the center, for (c) $\gamma_{\max}=0.1$ and (d) $\gamma_{\max}=0.5$.
    Results are shown for three cases: time-independent dissipation ($\gamma = \gamma_{\max}$), time-dependent dissipation $\gamma(t)$, and the random Lindblad protocol (random operators and sites).}
    \label{fig:combined_lindblad}
\end{figure}

In the second protocol, we fix $\gamma$ to a constant value but randomise both the operator type (choosing randomly from $X$, $Y$, or $Z$ Pauli operators) and the sites at which the Lindblad operators act. For each simulation run, two sites are selected at random, and each is assigned a randomly chosen Pauli operator as the Lindblad operator. Figure \ref{fig:combined_lindblad} displays the time evolution of the local magnetisation $\langle \sigma^z_{i}(t)\rangle$ at the center of the chain (top row) and the connected correlation function $C^z_{0,N/2}(t)$ (bottom row), for both protocols and for two values $\gamma_{\max} = 0.1$ and $\gamma_{\max} = 0.5$. For the correlations, we focus on the distance $d$ between site $i=0$ and the center of the chain, but note that the observed behaviour is representative for other values of $d$ as well.

\begin{figure}[!ht]
    \centering
    \includegraphics[width=0.55\linewidth]{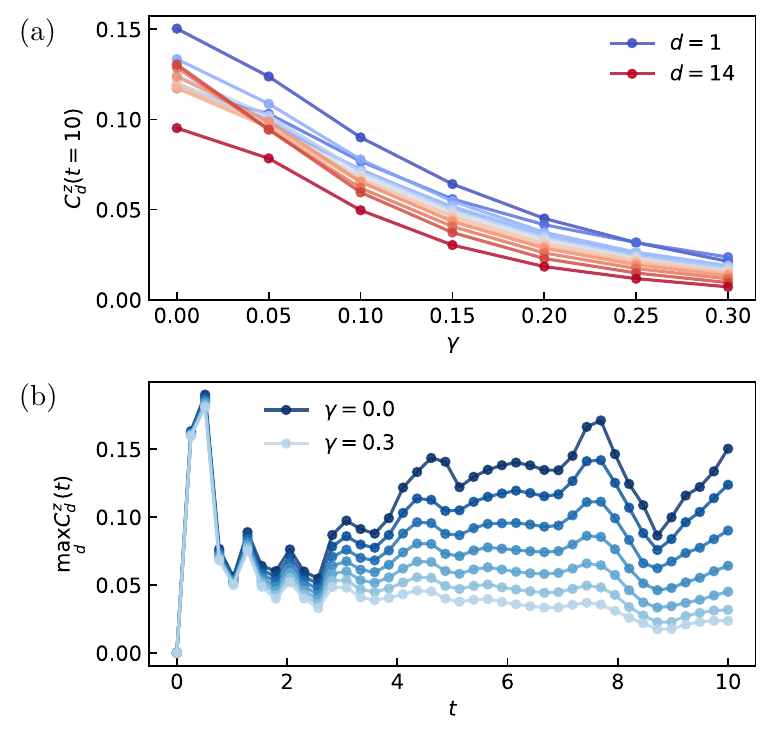}
    \caption{Spin–$Z$ correlations in a dissipative spin chain of with $N=25$ sites as a function of $\gamma$. 
    (a) Two‐point correlation $C^z_d(t=T)$ at the final time $T=10$, plotted for separations $d=1,\dots,N-1$ (curves range from deep red at $d=1$ to deep blue at $d=N-1$); labels indicate the shortest and longest distances.
    (b) Maximum correlation across distances as a function of $t$. Similarly, $\gamma \in [0, 0.3]$ with step size of $0.05$.}
    \label{fig:correlations_chaos}
\end{figure}

\begin{figure}[!ht]
    \centering
    \includegraphics[width=0.55\linewidth]{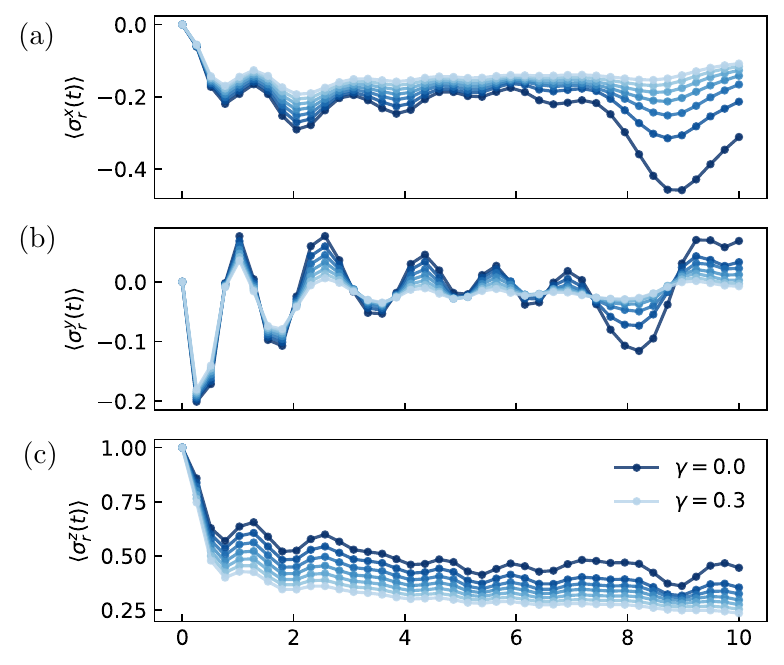}
    \caption{Time evolution under Lindbladian dynamics of spin observables for $S$ of size $N = 25$ under increasing dissipation rates $\gamma \in [0,0.3]$ with step size of $0.05$. (a–c) Expectation values of the local spin operators $\langle \sigma^x_r(t) \rangle$, $\langle \sigma^y_r(t) \rangle$, and $\langle \sigma^z_r(t) \rangle$, respectively.}
    \label{fig:pauli_expectation_chaos}
\end{figure}

These results highlight how both the temporal profile and spatial structure of the dissipation channel can influence relaxation and steady-state properties of local observables in open quantum systems. Notably, we find that the time-dependence of $\gamma$ in the first protocol does not lead to significant differences in either local magnetisation or correlations when compared to the corresponding time-independent case with $\gamma = \gamma_\mathrm{max}$. This is expected, as the overall effect of $\gamma$ can be understood as accumulating over time, but never exceeding $\gamma_\mathrm{max}$, so that both the relaxation and steady-state behaviour are largely determined by $\gamma_\mathrm{max}$. The correlations show a similar insensitivity to the time dependence of the dissipation, being primarily controlled by $\gamma$ itself. This behaviour persists in different choices of $d$.

In contrast, the second protocol, where Lindblad operators act at random sites with randomly chosen Pauli types, produces more noticeable and consistent differences in the observable dynamics. Here, the randomness in the spatial and operator structure of the dissipation can substantially alter the system's relaxation pathways, leading to a richer variety of dynamical behaviours in the observables, depending on the particular realisation of the dissipative channel. As discussed in the main text, in the context of ground-state preparation through dissipative protocols, this suggests that a time-dependent, randomised application of jump operators could accelerate convergence and enable more efficient state preparation.

\begin{figure}[!t]
    \centering
    \includegraphics[width=0.55\linewidth]{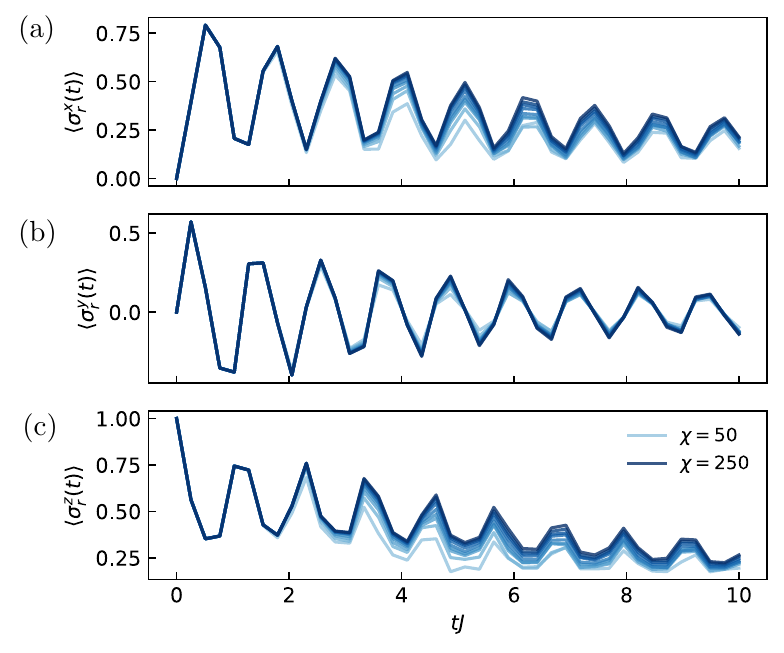}
    \caption{Time evolution of spin observables for a system of size $N=25 $ for increasing bond dimension $\chi$. (a-c) Expectation values of local spin operators $\langle \sigma^x_r(t) \rangle$, $\langle \sigma^y_r(t) \rangle$, and $\langle \sigma^z_r(t) \rangle$, respectively, under Lindbladian dynamics for $\chi\in [50, 250]$, with increasing values shown in progressively darker shades of blue.}
    \label{fig:obs_bond_dimension}
\end{figure}

\section{Tensor network simulation results for chaotic systems}
\label{sec:app_tensor_network}

Here, we present additional results from TN simulations. Although the main text focuses on the Heisenberg model with oscillatory behaviour in its Pauli observables, in Fig.~\ref{fig:pauli_expectation_chaos} we show the case for $h_x=0.5$, $h_z=-1.05$, corresponding to a chaotic regime.  
Here, the coherent oscillations present at weak dissipation are rapidly suppressed as $\gamma$ increases, leading to faster relaxation towards steady state. This behaviour illustrates how dissipation tends to erase the signatures of chaos in local observables, effectively driving the system into featureless dynamics at long times. Evidently, the correlation length exhibits the same trend, becoming increasingly damped with $\gamma$---more noticeably than in the case studied in the main text---and with total evolution time, as shown in Fig.~\ref{fig:correlations_chaos}.  
This confirms that dissipation not only suppresses temporal oscillations but also washes out spatial correlations across the chain, reinforcing the picture that strong coupling to the environment reduces the effective complexity of the dynamics.

On a different note, in this work we set the bond dimension to $\chi=250$ as the ground truth. In Fig.~\ref{fig:obs_bond_dimension} we show that this choice is well justified, with the observable dynamics clearly converging as $\chi$ increases.  
The convergence is particularly evident at intermediate times, where entanglement growth is more pronounced, yet the difference between $\chi=200$ and $\chi=250$ remains negligible. This validates our numerical accuracy and ensures that all qualitative conclusions drawn in the main text are robust with respect to truncation errors.


\end{document}